\documentclass[11pt]{article} %
\usepackage[ascii]{inputenc}

\usepackage{multicol} \usepackage{picinpar}

\usepackage{epsfig}

\usepackage{color}

\usepackage{ifthen} \usepackage{latexsym} \usepackage{amsfonts}
\usepackage{amsmath} \usepackage{amssymb}
\usepackage{pifont}

\def\desclabel#1{\bf #1\hfil} \def\desc{\list{}{%
    \setlength{\leftmargin}{0pt}
    \labelwidth= \leftmargin \advance \labelwidth by -\labelsep \let
    \makelabel=\desclabel}}

\def\descHACKlabel#1{\bf #1\hfil} \def\descHACK{\list{}{%
    \setlength{\leftmargin}{0pt}
    \labelwidth= \leftmargin \advance \labelwidth by -\labelsep \let
    \makelabel=\descHACKlabel}} 

\newcounter{extremeleftlistcounter} %
{\begin{list}{\arabic{extremeleftlistcounter}~~~}{\usecounter{extremeleftlistcounter}%
      \setlength{\labelsep}{0pt}\setlength{\leftmargin}{0pt}%
      \setlength{\labelwidth}{0pt}\setlength{\listparindent}{0pt}}}%
  {\end{list}}

\newcounter{leftlistcounter} %
{\begin{list}{\arabic{leftlistcounter}~~~}{\usecounter{leftlistcounter}%
      \setlength{\labelsep}{0pt}\setlength{\leftmargin}{15pt}%
      \setlength{\labelwidth}{15pt}\setlength{\listparindent}{0pt}}}%
  {\end{list}}

\newcommand{\hugeDebug}{false} \ifthenelse{\equal{\hugeDebug}{false}}{
   \pagestyle{plain}
  \setlength{\oddsidemargin}{0.25in}
  \setlength{\evensidemargin}{\oddsidemargin}
  \setlength{\textwidth}{6in} \setlength{\textheight}{8in}
  \setlength{\topmargin}{-0.0in} } {
   \pagestyle{headings}
  \setlength{\oddsidemargin}{-0.4in}
  \setlength{\evensidemargin}{\oddsidemargin}
  \setlength{\textwidth}{5.3in} \setlength{\textheight}{6in}
  \setlength{\textheight}{6.25in} \setlength{\topmargin}{-0.0in} }

    \newlength{\filength}
 \newsavebox{\gcbox}
\sbox{\gcbox}{\framebox[\filength]{\rule{0ex}{2ex}}}

\newtheorem{theorem}{Theorem}[section]
\newtheorem{corollary}[theorem]{Corollary}

\newcommand\qedblob{\ding{113}} \def\literalqed{{\
    \nolinebreak\hfill\mbox{\qedblob\quad}}}

 \newtheorem{lemma}[theorem]{Lemma}

\newtheorem{observation}[theorem]{Observation}

 \showhyphens{}

\newcount\hour \newcount\minutes \hour=\time \divide\hour by 60
\minutes=\hour \multiply\minutes by -60 \advance\minutes by \time
\def\mmmddyyyy{\ifcase\month\or Jan\or Feb\or Mar\or Apr\or May\or
  Jun\or Jul\or Aug\or Sep\or Oct\or Nov\or Dec\fi \space\number\day,
  \number\year} \def\hhmm{\ifnum\hour<10 0\fi\number\hour :%
  \ifnum\minutes<10 0\fi\number\minutes}

\makeatletter
\def\@citex[#1]#2{\if@filesw\immediate\write\@auxout{\string\citation{#2}}\fi
  \def\@citea{}\@cite{\@for\@citeb:=#2\do
    {\@citea\def\@citea{,\linebreak[0]}\@ifundefined {b@\@citeb}{{\bf
          ?}\@warning {Citation `\@citeb' on page \thepage \space
          undefined}}%
      \hbox{\csname b@\@citeb\endcsname}}}{#1}}
\newcommand{\singlespacing}{\let\CS=
  \@currsize\renewcommand{\baselinestretch}{1}\tiny\CS}
\newcommand{\singlespacingplus}{\let\CS=
  \@currsize\renewcommand{\baselinestretch}{1.25}\tiny\CS}
\newcommand{\doublespacing}{\let\CS=
  \@currsize\renewcommand{\baselinestretch}{1.75}\tiny\CS}
\newcommand{\extradoublespacing}{\let\CS=
  \@currsize\renewcommand{\baselinestretch}{1.9}\tiny\CS}
\newcommand{\nicenicespacing}{\let\CS=
  \@currsize\renewcommand{\baselinestretch}{1.9}\tiny\CS}
\newcommand{\draftspacing}{\let\CS=
  \@currsize\renewcommand{\baselinestretch}{2.0}\tiny\CS}
\newcommand{\hugedraftspacing}{\let\CS=
  \@currsize\renewcommand{\baselinestretch}{2.4}\tiny\CS}
\newcommand{\niceonespacing}{\let\CS=\@currsize\renewcommand{\baselinestretch}{1.1}\tiny\CS}
\newcommand{\nicetwospacing}{\let\CS=\@currsize\renewcommand{\baselinestretch}{1.2}\tiny\CS}
\newcommand{\nicethreespacing}{\let\CS=\@currsize\renewcommand{\baselinestretch}{1.3}\tiny\CS}
\newcommand{\singlespacingplusplus}{\let\CS=\@currsize\renewcommand{\baselinestretch}{1.35}\tiny\CS}
\newcommand{\nicefourspacing}{\let\CS=\@currsize\renewcommand{\baselinestretch}{1.4}\tiny\CS}
\newcommand{\nicefivespacing}{\let\CS=\@currsize\renewcommand{\baselinestretch}{1.5}\tiny\CS}
\newcommand{\nicesixspacing}{\let\CS=\@currsize\renewcommand{\baselinestretch}{1.6}\tiny\CS}
\newcommand{\nicesevenspacing}{\let\CS=\@currsize\renewcommand{\baselinestretch}{1.7}\tiny\CS}
\newcommand{\niceeightspacing}{\let\CS=\@currsize\renewcommand{\baselinestretch}{1.8}\tiny\CS}
\newcommand{\niceninespacing}{\let\CS=\@currsize\renewcommand{\baselinestretch}{1.9}\tiny\CS}
\makeatother%

\newcommand{\naturalnumber}{\ensuremath{{ \mathbb{N} }}}


 \newcommand{\sat}{{\rm SAT}}

\newcommand{\p}{{\rm P}} \newcommand{\littlep}{{\rm p}}
 
\newcommand{\npcc}{{\rm NPC}}
 \newcommand{\np}{{\rm NP}}

 \newcommand{\conp}{{\rm coNP}}

\newcommand{\npnp}{\ensuremath{\np^\np}}
\newcommand{\conpnpnp}{\ensuremath{\conp^{\np^\np}}}
\newcommand{\conpnp}{\ensuremath{\conp^\np}}

\newcommand{\pitwo}{\ensuremath{\conpnp}} %
\newcommand{\sigmatwo}{\ensuremath{\npnp}} %
\newcommand{\pithree}{\ensuremath{\conpnpnp}} %

\newenvironment{proofs}{\noindent{\bf
    Proof.}\hspace*{1em}}{\literalqed\bigskip}

\DeclareMathSymbol{\subsetneq}{\mathbin}{AMSb}{"28}
\DeclareMathSymbol{\supsetneq}{\mathbin}{AMSb}{"29}

\newcommand{\card}[1]{{ \mathopen\parallel {#1} \mathclose\parallel }}

\newcommand{\pair}[1]{\mathopen\langle{#1}\mathclose\rangle}

\newcommand{\manyone}{\ensuremath{\,\leq_{\rm m}^{{\littlep}}\,}}

\newcommand{\shortcite}[1]{\cite{#1}} %
\newcommand{\lhbox}[1]{\hbox{\,{#1}\,}} %
\newcommand{\vpair}[2]{\ensuremath{\left[{\rm{}#1} \atop
      {\rm{}#2}\right]}}
\newcommand{\vtriplestd}{\vtriple{PC}{RPC}{PV}}
\newcommand{\vtriple}[3]{\ensuremath{\left[ \substack{
        {\rm{}#1} \\[0.1ex]
        {\rm{}#2} \\[0.1ex]
        {\rm{}#3} } \right]}} \newcommand{\vpaircd}{{\vpair{C}{D}}}
\newcommand{\vpairad}{{\vpair{A}{D}}}
\newcommand{\vpaircv}{{\vpair{C}{V}}}
\newcommand{\vpairtetp}{{\vpair{TE}{TP}}}
\newcommand{\vpaircfmf}{{\vpair{CF}{MF}}} \newcommand{\probbf}{\rm}
\newcommand{\scontrol}[2]{{\probbf \mbox{#1}\hbox{-}\mbox{\rm{}#2}}}
\newcommand{\mcontrol}[2]{{\probbf
    \mbox{#1}\hbox{-}M{+}\mbox{\rm{}#2}}}
\newcommand{\escontrol}[1]{{\scontrol{\elec}{#1}}}
\newcommand{\mescontrol}[1]{{\escontrol{M{+}#1}}}

\newcommand{\scontrolcf}[2]{{\scontrol{#1}{{#2\hbox{-}CF}}}}
\newcommand{\scontrolmf}[2]{{\scontrol{#1}{#2\hbox{-}MF}}}

\newcommand{\scontrolmfrevoting}[2]{{\scontrol{#1}{#2\hbox{-}MF\hbox{-}revoting}}}

\newcommand{\escontrolcf}[1]{{\scontrolcf{\elec}{#1}}}
\newcommand{\escontrolmf}[1]{{\scontrolmf{\elec}{#1}}}

\newcommand{\escontrolmfrevoting}[1]{{\scontrolmfrevoting{\elec}{#1}}}

\newcommand{\ducm}[1]{\ensuremath{\mbox{\rm{}#1}\hbox{-}\allowbreak\mbox{\rm{}DUCM}}}
\newcommand{\cucm}[1]{\ensuremath{\mbox{\rm{}#1}\hbox{-}\allowbreak\mbox{\rm{}CUCM}}}

\newcommand{\elec}{\electionsystem}
\newcommand{\elecprime}{\ensuremath{\electionsystem'}}
\newcommand{\electionsystem}{\cale}
\newcommand{\cale}{\ensuremath{\cal E}}

\newcommand{\score}[1]{{{\mbox{\it{score}}(#1)}}}
\newcommand{\scoresub}[2]{{{\mbox{\it{score}}_{#1}(#2)}}}

\newcommand{\mymin}[1]{{{\mbox{\rm{min}}{(#1)}}}}

\newcommand{\qbfsigma}[1]{\ensuremath{{\rm QBF}_{#1}}}
\newcommand{\qbfpi}[1]{\ensuremath{\widetilde{{\rm QBF}}_{#1}}}

\newcommand{\bigcdot}{\ensuremath{\,\! \boldsymbol{\cdot}\,\!}}

\title{Control in the Presence of Manipulators: Cooperative and
  Competitive Cases\thanks{Supported in part by NSF grants
     CCF-0915792, CCF-1101452, CCF-1101479, and
     NSF Graduate Fellowship DGE-1102937.  Earlier 
      versions of this paper~\cite{fit-hem-hem:c:control-manipulation,fit-hem-hem:c:control-manipulation-comsoc} 
  were presented at 
  the 
  Twenty-Third International Joint Conference on Artificial Intelligence
 and
  the
  Fifth International Workshop on Computational Social 
 Choice.}}

\author{Zack Fitzsimmons \\
  College of Computing and Inf.\ Sciences\\
  Rochester Inst.\ of Technology \\
  Rochester, NY 14623 \and
  Edith Hemaspaandra \\
  Dept.~of Computer Science\\
  Rochester Inst.\ of Technology \\
  Rochester, NY 14623 \and
  Lane A. Hemaspaandra \\
  Dept.~of Computer Science\\
  University of Rochester \\
  Rochester, NY 14627 }

\date{August 2, 2013; revised March 3, 2017 and June 8, 2017}

\begin{document}
\sloppy

\maketitle

\begin{abstract}
  Control and manipulation are two of the most studied types of
  attacks on elections. In this paper, we study the complexity of
  control attacks on elections in which there are manipulators. We
  study both the case where the ``chair'' who is seeking to control
  the election is allied with the manipulators, and the case where the
  manipulators seek to thwart the chair. In the latter case, we see
  that the order of play substantially influences the complexity. We
  prove upper bounds, holding over every election system with a
  polynomial-time winner problem, for all standard control cases, and
  some of these bounds are at the second or third level of the
  polynomial hierarchy, and we provide matching lower bounds to prove
  these tight. Nonetheless, for important natural systems the
  complexity can be much lower. We prove that for
  approval and plurality elections, the complexity of even competitive
  clashes between a controller and manipulators falls far below those
  high bounds, even as low as polynomial time.
  Yet for a Borda-voting case we show that such clashes raise the
  complexity unless $\np = \conp$.
\end{abstract}

\section{Introduction}
Elections are an important tool in reaching decisions, in both human
and online settings.  
Regarding online settings, elections have been proposed
in such varied, multiagent-systems settings as 
planning, recommender systems/collaborative filtering, and web spam 
reduction~\cite{eph-ros:j:multiagent-planning,gho-her-mun-sen:c:voting-for-movies,gil-hor-pen:c:collaborative-filtering,fag-kum-siv:c:similarity-search}.
With the 
growing importance of the online
world and multiagent systems, the use of elections in computer-based
settings will but increase.  Unfortunately, given the relentless
growth in the power of computers, it is natural to worry that
computers will also be increasingly brought to bear
in planning manipulative attacks on elections.
Indeed, this is one of the central concerns of the relatively 
young multiagent systems subarea known as computational
social choice~\cite{che-end-lan-mau:c:polsci-intro,bra-con-end:b:comsoc}.

The two most computationally studied types of attacks on elections are
known as ``control'' and ``manipulation.''  Both were introduced by
Bartholdi, Tovey, and
Trick~\shortcite{bar-tov-tri:j:manipulating,bar-tov-tri:j:control}.
In control, an agent, usually referred to as ``the chair,'' tries to
make a given candidate win by adding/deleting/partitioning voters or
candidates.  In manipulation, some nonmanipulative voters and a
coalition of manipulative voters vote under some election system, and
the manipulative voters seek to make a given candidate win.  

There is
a broad literature on the computational complexity of control, and on
the computational complexity of manipulation.
However, the present 
paper considers \emph{control attacks against elections that contain
manipulators}.  We consider both the cooperative and the competitive
cases.

In the cooperative case, the chair is allied with the manipulative
coalition.  For example, perhaps during a CS department's hiring, the
department chair, who happens to also be the senior member of the
systems group, is mounting a control by partition of voters attack (in
which he or she is dividing the faculty into two subcommittees, one to
decide which candidates are strong enough teachers to merit further
consideration, and one to decide which candidates are strong enough
researchers to merit further consideration), and also is able to
directly control the votes of every one of his or her fellow members
of the department's systems faculty.  The chair's goal is to make some
particular candidate, perhaps Dr.\ I. M. Systems, be the one chosen
for hiring.

In the even more interesting %
competitive case, which can be thought of in a certain sense as
\emph{control versus manipulation}, we will assume that the manipulative
coalition's goal is to keep the chair from achieving the chair's goal.
For the competitive case, we will look at the case where the chair acts
before the manipulators, and at the case where the manipulators act
before the chair.  For control attacks by so-called partition, in
which there is a two-round election, we will consider the case where the
manipulators can change their votes in the second round, and the case
where the manipulators cannot change their votes in the second round.

Our main contributions are as follows.

\begin{itemize}
\item Building on the existing notions of control and manipulation, we give
natural definitions that capture our cooperative and competitive
notions as problems whose computational complexity can be studied,  and we
note how existing hardness results for control and manipulation are,
or are not, inherited by our problems.

\item We prove upper bounds on our
problems.  For the competitive case, these are as high as
$\npnp$, $\conpnp$, and $\conpnpnp$, with the notable exception of
the case of deleting voters in the ``chair-first'' setting, which is in
coDP, i.e., it is the union of an \np\ set and a \conp\ set.

\item Despite how high those upper bounds are, we show %
that there are election systems (having polynomial-time winner problems) for
which most of those high bounds have matching lower bounds, yielding
completeness for those classes.

\item For the important election systems approval, Condorcet, and plurality,
we show that the complexity of control in the presence of
manipulators, whether cooperative or competitive, can be much lower than %
those upper bounds,
even falling as low as polynomial time.
Many of the proofs of these cases involve 
novel approaches---%
approaches very 
different 
than
those used in 
the case of control without manipulators
(see, e.g., the proof of Theorem~\ref{thm:con-m+ccpc}).

\item 
We obtain results, for election systems satisfying
versions---called WARP and unique-WARP---of 
the weak axiom of revealed preferences,
on the complexity of control by runoff partitioning of
candidates.

The general theme of those results is that the 
combinatorial explosion that causes many
partition-related candidate-control 
problems to be NP-complete can never exist for 
election systems that satisfy certain nice properties,
such as WARP and unique-WARP\@.  In particular, we will show that 
such properties can change the challenge facing the 
chair (of a control problem) from that of needing to worry about 
every partition to just that of 
checking one very simple partition.  From 
this, polynomial-time control algorithms immediately follow, as we will see.

The reason that this is interesting is that it is not applying just to one 
particular system, but rather is noting that some nice behavioral 
properties themselves 
ensure the simplicity of certain candidate-partition control problems for 
\emph{all} systems having the properties.

\item We also obtain cases, for veto (Theorem~\ref{t:veto-surprise}) and
Borda (Theorem~\ref{t:borda-surprise}) elections, where competitive
control-plus-manipulation is variously easier or harder than one might
expect from the separate control and manipulation cases.

\end{itemize}

\section{Related Work}
The idea of enhancing control with manipulative voters has been
mentioned in the literature, namely, in a paragraph of
\cite{fal-hem-hem:j:multiprong}.  That paper cooperatively integrated
with control, to a certain extent, a different attack type known as
bribery \cite{fal-hem-hem:j:bribery}.  In that paper's conclusions and
open directions, there is a paragraph suggesting that manipulation
could and should also be integrated into that paper's ``multiprong
setting,'' and commending such future study to interested readers.
That paragraph was certainly influential in our choice of this
direction.  However, it is speaking just of the cooperative case, and
provides no results on this since it is suggesting a direction for
study.

The lovely line of work about ``possible winners''
\cite{kon-lan:c:incomplete-prefs} in the context of adding candidates
might at first seem to be merging manipulation and control.  We refer
to the line of work explored in
\cite{che-lan-mau-mon:cOutByJour:possible-winners-adding,bai-roo-rot:c:two-variants-possible-winner,lan-mon-xia:c:new-alternatives-new-results,che-lan-mau-mon-xia:j:possible-winners-adding-welcome}.
That work considers an election with an initial set of candidates,
over which all the voters have complete preferences, and a set of
additional candidates over which the voters initially have no
preferences, and asks whether, if the entire set of additional
candidates is added, there is some way of extending the initial linear
orders to now be over all the candidates, in such a way that a
particular initial candidate becomes a winner of the election.
Although on its surface this might feel like a cross between
manipulation and control by adding candidates, in fact, in this
interesting problem there is no actual choice
regarding the addition of candidates; all are simply added.  Thus this
problem is a generalization of manipulation (as the papers note), that
happens to be done in a setting that involves adding candidates.  It
is not a generalization of control by adding, or even so-called
unlimited adding, of candidates, as in those the chair must choose
what collection of candidates to add.  In short, unlike control and
unlike this paper, there is no existentially quantified action by a
chair.  (An interesting recent paper of Baumeister et
al.~\shortcite{bau-roo-rot-sch-xia:c:possible-weight-winners} uses the
term possible winner in a new, different way, to speak of weights
rather than preferences initially being partially unset.  That
particular paper's question, as that paper notes, can be seen as a
generalization of control by adding and deleting voters.  However,
their notion is not a generalization of manipulation.)

The present paper does combine control and manipulation, with both
those playing active---and sometimes opposing---roles.  Manipulation
alone has been extensively studied in a huge number of papers,
starting with the seminal paper of \cite{bar-tov-tri:j:manipulating}
(see also \cite{bar-orl:j:polsci:strategic-voting}), which covered the
constructive case.  The destructive cases (i.e., those where the goal
is to keep a particular candidate from winning) are due to
\cite{con-lan-san:j:when-hard-to-manipulate}.  Control alone has been
extensively studied in many papers, with the seminal paper being
\cite{bar-tov-tri:j:control}, which covered the constructive case.
The destructive cases were first studied in
\cite{hem-hem-rot:j:destructive-control}.  There has been quite a bit
of work on finding systems for which conducting various types of
manipulation is hard, or for which conducting most types of control
attacks is hard, see, e.g.,
\cite{erd-now-rot:j:sp-av,fal-hem-hem-rot:j:llull,hem-hem-rot:j:hybrid,men:j:range-voting,men-sin:c:schulze,par-xia:c:ranked-pairs,erd-fel-rot-sch:j:control-in-bucklin-and-fallback-voting}
or the
surveys~\cite{fal-hem-hem-rot:b:richer,fal-hem-hem:j:cacm-survey,con-wal:b:barriers-to-manipulation-in-voting,fal-rot:b:handbook-comsoc-control-and-bribery}.

In the present paper, we will see that who goes first, the chair or the
manipulators, is important in determining what complexity upper bounds
apply.  Order has also been seen to be important in the study of
so-called online control
attacks~\cite{hem-hem-rot:c:online-voter-control,hem-hem-rot:c:online-candidate-control},
and of online manipulation
attacks~\cite{hem-hem-rot:j:online-manipulation}.  However,
the papers just mentioned are separately about control, and about
manipulation.  In contrast we are mostly interested in when both are
occurring, and especially when the two attacks are in conflict with
each other.

The present paper also looks at how revoting affects the complexity of
elections that involve both control and manipulation.  It is important
to mention that, for the case of just manipulation,
\cite{nar-wal:c:two-stage-comsoc,nar-wal:c:two-stage-aamas} 
(see also~\cite{fit-hem-hem:j:xthenx}) 
have recently discussed revoting, and give
an example that shows that revoting can sometimes be a valuable tool
for the manipulator.

\section{Preliminaries}
An election (a.k.a.\ a social choice correspondence) maps from a
finite candidate set $C$ and a finite vote collection $V$ to a set, $W
\subseteq C$, called the
winner(s)~\cite{ley-sho:b:multiagent-systems}.
Candidates each have a corresponding name, and these names play an
important role in some of our results.
Voters come without
names, and the votes are input as a list, i.e., as ballots.  For
approval elections, each ballot is a length-$\card{C}$ 0-1 vector
indicating whether each candidate is disapproved or approved.  The
candidate getting the most approvals is the winner (or winners if
candidates tie for most).  For all other systems we discuss, each
ballot is a tie-free linear ordering of the candidates.  For plurality
elections, each voter gives one point to his or her top choice and
zero to the rest.  For veto elections, each voter gives zero points to
his or her bottom choice and one to the rest.  For Borda elections,
each voter gives zero points to his or her bottom choice, one point to
his or her next to bottom choice, and so on through giving
$\card{C}-1$ points to his or her top choice.  In the three systems
just mentioned, the winner is the candidate(s) who receives the most
points.  In a Condorcet election---\cite{bar-tov-tri:j:control} recast
the notion of a Condorcet winner \cite{con:b:condorcet-paradox} into
an election system of sorts, in this way, and used it as one of their
focus cases in their seminal control study---a candidate $p$ is a
winner exactly if for each other candidate $b$ it holds that strictly
more than half the votes cast prefer $p$ to $b$.  Unlike the systems
from earlier in this paragraph, Condorcet elections on some inputs may
have no winners.

An election system $\elec$ is said to have a p-time winner problem if
there is a polynomial-time algorithm that on input $C$, $V$, and $p\in
C$, determines whether $p$ is a winner under $\elec$ of the election
over $C$ with the votes being $V$.

We assume the reader is aware of the NP, $\conpnp$, $\npnp$, and
$\conpnpnp$ levels of the polynomial hierarchy (the ``exponentiation''
notation denotes oracle class, informally put, having unit-cost access
to a set of one's choice from the given
class)~\cite{mey-sto:c:reg-exp-needs-exp-space,sto:j:poly}.
DP is the class of languages that are the
difference of two languages in \np~\cite{pap-yan:j:dp}.
We assume that the reader is familiar with
many-one reductions (which here always means many-one polynomial-time
reductions).  As is standard, we use $\manyone$ to denote many-one
reductions.  There are far fewer completeness results for levels of
the hierarchy beyond NP, such as the abovementioned ones, than there
are for NP; a collection of and discussion of such results can be
found in \cite{sch-uma:j:PH-part-one,sch-uma:j:PH-part-two}.
Completeness and hardness here are always with respect to many-one
reductions.

For proofs of the cases of Theorem~\ref{t:theoretical-bounds} we reduce from
Quantified Boolean Formulas~(QBF) where formulas are restricted to $k$
alternating quantifiers where each quantifier quantifies over a list
of boolean variables.
The problem $\qbfsigma{k}$ is the case of
$k$ alternating quantifiers beginning with $\exists$ and similarly
$\qbfpi{k}$ is the case of $k$ alternating quantifiers beginning with
$\forall$; $\qbfsigma{2}$ is $\npnp$-hard,
$\qbfpi{2}$ is $\conpnp$-hard, and $\qbfpi{3}$
is $\conpnpnp$-hard~\cite{sto-mey:c:word-problems,wra:j:complete}.
In all our proofs using $\qbfsigma{k}$ or $\qbfpi{k}$ we assume
without loss of generality that the same number of variables are
bound to each quantifier.

Our hardness results are worst-case results.  However, it is known
that if there exists even one set that is hard for NP (and note that
all sets hard for $\conpnp$, $\npnp$, or $\conpnpnp$ are hard for NP)
and has a (deterministic) heuristic algorithm whose asymptotic error
rate is subexponential, then the polynomial hierarchy collapses.
See~\cite{hem-wil:j:heuristic-algorithms-correctness-frequency} for a
discussion of that, and an attempt to reconcile that with the fact
that in practice heuristics often do seem to do well, including for
some cases related to elections, see, e.g.,
\cite{wal:c:where-hard-veto}.

\subsection{Types of Electoral Control}

We now briefly define all standard control types.
For a more formal description
we refer the reader to the detailed definitions given in
\cite{fal-hem-hem-rot:j:llull}. Given as input
an election, $(C,V)$, a distinguished candidate $p\in C$, and an
integer $k \geq 0$, the constructive (respectively, destructive)
control by deleting voters---for short CCDV (respectively,
DCDV)---problem for an election system $\elec$ asks whether there is
some choice of at most $k$ votes such that if they are removed, $p$ is
a winner (respectively, is not a winner) of the given election under
$\elec$.  We are in the so-called nonunique-winner model, and so we
ask about making $p$ ``a winner'' rather than ``the one and only
winner,'' which is the so-called unique-winner model.\footnote{Many of
  our results also hold in the other model, but the nonunique-winner
  model is probably the better, more natural model on which to focus
  in general.}
Each of those problems has an adding voters~(AV)
analogue, in which one has a collection of registered voters,
and has a collection of ``unregistered'' voters,
and the question is whether there is some choice of at most $k$
voters from the collection of unregistered voters such that if they are added, the goal is met.
These types of control are motivated by issues ranging from voter
suppression to targeted phone calls to get-out-the-vote drives.  There
are the natural analogous types for adding and deleting candidates, AC
and DC (note: in the destructive control by deleting candidates
case---DCDC---deleting $p$ is not allowed
\cite{bar-tov-tri:j:control}).

The partition types are called runoff
partition of candidates~(RPC), partition of candidates~(PC),
and
partition of voters~(PV).
In each of the three partition control types, the input is just $(C,V)$ and $p \in C$, and
a two-stage election is performed. In RPC, the constructive (destructive) question is whether
there exists a partition of the candidates into $C_1$ and $C_2$ such that if the candidates
who survive at least one of the elections $(C_1,V)$ and $(C_2,V)$ move on to a runoff %
among just them with the collection of votes $V$, $p$ is (is not) a winner. 
(Though we write ``$V$'' for the voter set in each subelection, 
that implicitly means $V$ masked down just to the candidates at 
hand in the subelection; the analogous issue holds 
regarding the DC case; and in 
the AC case, the voters' preferences $V$ are over the set of all 
registered and unregistered candidates and are also similarly masked down 
when called upon.)
Here, there are two models for what ``survive'' means. In the ties eliminate~(TE) model,
to move forward one must uniquely win a first-round election; in the ties promote~(TP) model,
it suffices to be \emph{a}~winner of a first-round election.
The PC case is similar, but the winners of the election $(C_1,V)$ move on to a runoff with %
all the candidates in $C_2$.\footnote{Recent work by Hemaspaandra, Hemaspaandra, and Menton shows that in the nonunique
winner model two pairs of the standard control models collapse. Specifically, the models
of destructive control by partitioning candidates and destructive control by runoff partitioning
candidates, in each of the tie-breaking models~\cite{hem-hem-men:c:search-vs-decision}.}
In PV, we instead consider a partition of the collection of voters $V$ into
$V_1$ and $V_2$ where the runoff consists of the candidates that survive at
least one of the elections $(C,V_1)$ and $(C,V_2)$.

\subsection{Manipulation}
As to manipulation, the constructive (destructive) unweighted
coalitional manipulation CUCM (DUCM) problem under election system
$\elec$ has as input $(C,V)$, $p \in C$,
and a collection of manipulator voters,
and the question is whether there is some way
of setting the votes of the manipulative coalition
so that $p$ is (is not) a winner of the resulting election under
system $\elec$ with those votes and the nonmanipulative votes both
being cast.

\subsection{Control-plus-Manipulation}

Our model of allowing control in the presence of manipulators varies
the standard control definitions to allow some of the voters to be
manipulators, and thus to come in as blank slates.  We mention that
for AV, it is legal to have manipulators among the registered and/or
the unregistered votes.  For the cooperative cases, the question is
whether the chair can choose preferences for the manipulators such
that, along with using his or her legal control-decision ability for
that control type, $p$ can be made (precluded from being) a winner.
We denote these types by adding in an ``M+,'' e.g.,
plurality-M+CCAV\@.  For the competitive cases, we can look at the
case where the manipulative coalition sets its votes and then the
chair chooses a control action, and we call that MF for ``manipulators
first.''  Or we can have the chair control first and then the
manipulators set their votes, which we call CF for ``chair first.''
Since the manipulators seek to thwart the chair, the case
Borda-CCAV-MF, for example, asks whether under Borda, no matter how
the manipulative voters, moving first, set their votes, there will
exist some choice of at most $k$ unregistered voters that the chair
can add so that
$p$ is a winner.  For partition cases, we add
the string ``-revoting''
to indicate that after the first-round elections occur, the
manipulators can change their votes in the runoff. %
Notice that for a given control action the CF case is a subset of the
MF case, since if there exists a control action such that for all manipulations
the chair is successful, then the chair is successful with this same control action
when the manipulators go first.

Below we formally state the control
plus manipulation action of constructive control by deleting
voters (CCDV) for the collaborative (M+), chair-first (CF), and
manipulator-first (MF) cases.

\begin{description}
  \item[Name:] \elec-M{+}CCDV/\elec-CCDV-CF/\elec-CCDV-MF

  \item[Given:] An election $(C,V \cup W)$ (where $V$ and $W$ denote
  the nonmanipulative and manipulative voters respectively),
  a preferred candidate $p \in C$,
  and a delete limit $k \in \naturalnumber$.
  
  \item[Question (M+):] Does there exist a subcollection $V'
  \subseteq (V \cup W)$ such that $\|V'\| \le k$, and a way to set the
  votes of the manipulators, such that $p$ is a winner of $(C,(V \cup W)-V')$
  under election system \elec?
  
  \item[Question (CF):] Does there exist a subcollection $V'
  \subseteq (V \cup W)$ such that $\|V'\| \le k$, so that regardless of how
  the manipulators set their votes, $p$ is a winner of $(C,(V \cup W)-V')$ under
  election system \elec?

  \item[Question (MF):] Regardless of how the manipulators set
  their votes, does there exist a subcollection $V' \subseteq (V \cup W)$ such
  that $\|V'\| \le k$, and $p$ is a winner of $(C,(V \cup W)-V')$ under election
  system \elec?
\end{description}

To allow many things to be spoken of compactly, we use
``stacked'' notation to indicate every possible string one gets by
reading across and taking one choice from each bracket one encounters
on one's path across the expression.  So, for example,
CC\vpairad{}V-\vpair{CF}{MF} refers to four control types, not just
two, and \mbox{{\vpaircd{}C \vpair{ \vpairad\vpaircv }{
      \vtriplestd\lhbox{-}\vpairtetp}}} refers to $2\times( 2\times2
\,+\, 3\times2) = 20$ control types.

Notice that for our competitive setting, we 
seem to be 
asymmetrically focusing on things from the perspective of the chair.
That is, regardless of whether the chair moves first or whether the
manipulators move first, our problems are always posed in terms of the
chair's constructive or destructive goal regarding the candidate $p$.
It would be natural to ask---and indeed, a conference referee asked us
to address the issue of---whether one can interestingly study the
competitive problem from the perspective of the manipulators rather
than that of the chair.  That is, in the 
MF case for example, one would ask whether the manipulators can
act so as to achieve or block victory for $p$, regardless of
the actions of the chair that follow.  And one could similarly look at the
CF case from the manipulators' perspective.
After all, in many
real-world settings, what one cares about may well be 
the perspective of the manipulators.  Thus being able to address this
issue would itself be an additional motivation for our paper.
Fortunately, in the competitive case---and this holds in both the
nonunique-winner model and the unique-winner model, and holds for all
types of constructive and destructive attacks discussed here---the
chair achieving his or her goal in the model where we view things from
the perspective of the chair is precisely the same as the manipulators
failing to meet their goal in the model where we view things from the
perspective of the manipulators.  This follows from the definitions.
Thus this paper is
implicitly handling the case of the manipulators' perspective:  
For all our competitive cases,
studying a constructive (respectively, destructive) attack problem
from the perspective of the manipulators is exactly the same as
studying 
the complement of %
the \emph{destructive} (respectively, \emph{constructive}) 
version of the
same problem in the model of this paper, that is, from the perspective
of the chair.
For example, the sets $\escontrolcf{DCAV}
\text{-ManipulatorFocus}$
and
$\overline{\escontrolcf{CCAV}
\text{-ChairFocus}}$
are the same on all syntactically legal inputs
(and they will of course differ on all syntactically illegal inputs).
(We will not use ``focus'' suffixes in this paper except in the previous
sentence, since 
in this paper our all our problems will implicitly be 
``-ChairFocus.'')
We caution that the above discussion should not 
be interpreted as saying that the constructive and destructive 
problems are each other's opposites.  That is not true, although 
there is a partial connection between these cases,
see 
the discussion 
in footnote~5 of~\cite{hem-hem-rot:j:destructive-control}.

\section{Results}

\subsection{Inheritance}\label{ss:inheritance}
Each control type many-one reduces to each of its cooperative and to
each of its competitive control-plus-manipulation variants, because
for
those variants the zero-manipulator cases degenerate to the pure
control case.  For example,
$\scontrol{\elec}{CCDV} \manyone \mescontrol{CCDV}$ and
$\scontrol{\elec}{CCDV} \manyone \escontrolmf{CCDV}$.  In particular,
NP-hardness results for control inherit upward to each related
cooperative and competitive case.

For manipulation, the inheritance behavior is not as broad, since
partition control cannot necessarily be ``canceled out'' by setting a
parameter to zero, as partition doesn't even have a numerical
parameter.  Nonpartition control types do display inheritance, but for
the competitive cases there is some ``flipping'' of the type of
control and the set involved.  For each constructive (respectively,
destructive) control type regarding adding or deleting candidates or
voters, destructive (respectively, constructive) manipulation many-one
reduces to the complement of the set capturing the competitive case of
the constructive (respectively, destructive) control type combined
with manipulation.  For example, $\cucm{\elec} \manyone
\overline{\escontrolcf{DCAC}}$ and $\ducm{\elec} \manyone
\overline{\escontrolmf{CCDV}}$.
For the cooperative cases there is no
``flipping.''  For each constructive or destructive control type
regarding adding or deleting candidates or voters, manipulation
many-one reduces to the cooperative case of that control type combined
with manipulation.  For example, $\cucm{\elec} \manyone
\mescontrol{CCAC}$ and $\ducm{\elec} \manyone \mescontrol{DCAC}$.

\subsection{General Upper Bounds and Matching Lower Bounds}
\label{sec:upper-general-lower}

\begin{table*}[!t]
  \scriptsize %
  \centering
  \begin{tabular}{c | c c c c}
    Problem & CF & CF-revoting & MF & MF-revoting\\ \hline 
    \\ [-2.0ex]
    $\elec\lhbox{-}$\vpaircd{}C\vpairad\vpaircv & $\npnp$ (coDP for DV) & N/A & ${\conpnp}$  & N/A \\
    [+2.0ex]
    $\elec\lhbox{-}$\vpaircd{}C{}\vtriplestd\lhbox{-}\vpairtetp
    & $\npnp$ & $\npnp$ & $\conpnp$ & \begin{tabular}{c} $\conpnp$ \ \ \ (TE) \\ $\conpnpnp$ (TP)\end{tabular}
  \end{tabular}
  \caption{Upper Bounds\label{t:upper-general}.  (N/A means not applicable.)}
\end{table*}

For election systems with p-time winner problems, all the cooperative
cases clearly have NP upper bounds.  But the upper bounds for the
competitive cases are far higher, falling in the second and third
levels of the polynomial hierarchy, as described by the following
theorem.

\begin{theorem}\label{t:upper-bounds}
  For each election system $\elec$ having a p-time winner problem, the
  bounds of Table~\ref{t:upper-general} hold.\footnote{Where the table
    says N/A---not applicable---the nonrevoting bounds just to the
    left of the box technically still hold; we say N/A simply to be
    clear that revoting cannot even take place in nonpartition cases,
    since there is no second round.}
\end{theorem}

Although the table's upper bounds clearly follow from the structure of
the problems
(only for the coDP cases is this nontrivial, see Theorem~\ref{CDV-CF:coDP}), the bounds are very high.  Can they be improved by some
cleverer approach?  Or are there systems with p-time winner problems
that show the bounds to be tight?  The following result establishes
that the latter holds; each of the cells in the table is tight for at
least some cases.

\begin{theorem}\label{t:theoretical-bounds}
  \begin{enumerate}
  \item For each of the eight problems on the top line of
    Table~\ref{t:upper-general}, and each of the columns on that line,
    there exists an election system $\elec$, which has a p-time winner
    problem,
    for which the named problem is complete for the named complexity
    class.\footnote{The CCDV-CF and DCDV-CF cases were incorrectly classified
    as \npnp\ in early versions~\cite{fit-hem-hem:c:control-manipulation,fit-hem-hem:tOutByConfWITHVERSION:control-manipulation,fit-hem-hem:c:control-manipulation-comsoc}.}

  \item For each of {\rm{}CCPV-TP} and {\rm{}CCPV-TE}, and each of the
    {\rm{}CF}, {\rm CF-revoting}, and {\rm MF} columns of
    Table~\ref{t:upper-general}, and each of the columns on that line, 
    there exists an election system $\elec$, which has a p-time winner
    problem,
    for which the named problem is complete for the named complexity
    class.
  \item There exists an election system $\elec$, which has a p-time
    winner problem, for which {\rm CCPV-TP-MF-revoting} is
    $\conpnpnp$-complete, and there exists an election system $\elec$,
    which has a p-time winner problem, for which {\rm{}CCPV-TE-MF-revoting}
    is $\conpnp$-complete.
  \end{enumerate}
\end{theorem}

The above result
says that the upper bounds are not needlessly high.  They are truly
needed, at least for some systems.  However, the constructions proving
the lower bounds are artificial and the construction involving the
third level of the polynomial hierarchy is lengthy and
difficult.\footnote{The third-level case has to overcome the specific,
  and as far as we know new, worry that in the second round, the
  first-round vote of the manipulators is no longer available.  Yet
  in a ``$\forall\exists\forall$'' context (which is the quantifier
  structure that models $\conpnpnp$), a particular existential choice
  has to handle only a particular value of the first $\forall$.  So to
  make the construction work, we need to in some sense have the
  first-round votes, which are no longer available, still cast a clear
  and usable shadow forward into the second round, at least in certain
  cases in the image of the reduction.  We achieve this, in particular
  by shaping the election system itself carefully to help realize this
  unusual effect.  Otherwise, we would not be capturing the right
  quantifier structure.}
  In particular, this leaves completely open
the possibility that for particular, important real-world systems,
even the competitive cases may be far simpler than those bounds
suggest.  In the coming section, we will see that indeed for some of the
most important real-world systems, even in the presence of
manipulators, the control problem is just as computationally easy as
when there are no manipulators.

We now present the proof of the CCAC-CF case
of Theorem~\ref{t:theoretical-bounds}, which
illustrates the general arguments used in the proof of this theorem.
The proofs of the other cases can be found in
Appendix~\ref{app:upperbound}.

\begin{theorem}\label{CCAC-CF:complete}
There exists an election system, \elec, with a p-time winner problem,
such that \escontrolcf{CCAC} is \sigmatwo-complete.\end{theorem}

\begin{proofs} 
Let \elec\ be defined in the following way. Given an election $(C,V)$,
if $\|V\| = 1$, $\|C\| \ge 1$ and the candidates in $C$ listed in increasing
lexicographic order %
are $c_0, c_1, \dots, c_{\ell}$, and $c_0$ encodes %
a boolean formula $\psi(x_1, \dots, x_{2\ell})$, then do the following.
For each $i$, $1 \le i \le \ell$, set $x_i$ to true if the lowest-order
bit of $c_i$ %
is 1 and otherwise set $x_i$ to false.
For each $i$, $1 \le i \le \ell$, set $x_{\ell+i}$ to true if the voter
states $c_{i} > c_0$ and otherwise set $x_{\ell+ i}$ to false.
If this is a satisfying assignment for $\psi$ then everyone wins.
In all other cases everyone loses.
That completes the specification of \elec.

Clearly \elec\ has a p-time winner problem, and by
Theorem~\ref{t:upper-bounds} we know that \escontrolcf{CCAC}
is in \sigmatwo. So what is left is to show that \escontrolcf{CCAC}
is \sigmatwo-hard.
    
Let $(\exists x_1, \ldots, x_\ell)(\forall x_{\ell+1},\ldots, x_{2\ell})
[\psi(x_1,\ldots,x_{2\ell})]$ be an instance of $\qbfsigma{2}$.
We construct an instance of \escontrolcf{CCAC} in the following way.
Let the candidate set $C$ consist of $p$ %
encoding the boolean formula $\psi$,
and let there be zero nonmanipulators and one manipulator.
Let the set of unregistered candidates contain $\ell$ pairs where
for each $i$, $1 \leq i \leq \ell$,
there is a candidate %
$p \bigcdot i_{\rm binary} \bigcdot 0$ and a candidate
$p \bigcdot i_{\rm binary} \bigcdot 1$.
(where $\boldsymbol{\cdot}$ denotes
concatenation and $i_{\rm binary}$ denotes $i$ encoded in binary).
Let the add limit $k = 2\ell$.\footnote{We
set $k = 2\ell$ instead of the obvious choice of $\ell$ since
then the same proof can be used for the similar cases that appear in the
appendix,
and this also nicely handles the case of
``control by unlimited adding of candidates.''}

If $(\exists x_1, \ldots, x_\ell)(\forall x_{\ell+1},\ldots, x_{2\ell})
[\psi(x_1,\ldots,x_{2\ell})] \in \qbfsigma{2}$, fix an assignment
to $x_1, \ldots, x_\ell$ such that  
$(\forall x_{\ell+1},\ldots, x_{2\ell})
[\psi(x_1,\ldots,x_{2\ell})]$ is true. 
For each $i$, $1 \leq i \leq \ell$, the chair adds the candidate,
call it $c_i$, from the $i$th pair
whose last bit corresponds to the value of $x_i$ in this assignment. 
Note that %
$p, c_1, \ldots, c_\ell$ are in 
increasing lexicographic order.
Then no matter what assignment to
$x_{\ell + 1}, \ldots, x_{2\ell}$ is induced by the
manipulator's vote,
formula $\psi$
is satisfied and so $p$ will win.

Conversely, if the chair makes $p$ a winner,
then the chair adds exactly $\ell$ candidates
whose lowest-order bits give an assignment to
$x_1, \ldots, x_\ell$  such that
$(\forall x_{\ell+1},\ldots, x_{2\ell})
[\psi(x_1,\ldots,x_{2\ell})]$ is true.~\end{proofs}

\subsection{Specific Systems}

Plurality is certainly the most important of election systems, and approval
is also an important system.
Plurality, approval, and
Condorcet elections each have easy manipulation problems, and their complexity for
every standard control type is known~\cite{bar-tov-tri:j:control,hem-hem-rot:j:destructive-control}.
We display these known results in Table~\ref{tbl:cm:specific}.\footnote{It should be noted that the referenced table in~\cite{hem-hem-rot:j:destructive-control}
is focused on the unique-winner case, but by Observation~\ref{obs:winner-model} below these results carry over to the
nonunique-winner model (some of the
    cases were previously noted in Faliszewski, Hemaspaandra,
    and Hemaspaandra~\cite{fal-hem-hem:j:nearly-sp} and Hemaspaandra, Hemaspaandra, and
    Rothe~\cite{hem-hem-rot:c:online-voter-control}). Also, note that the ``AC'' line of the referenced
    table refers to so-called unlimited adding and
    (as is now standard) we use ``AC'' to refer to
    (limited) adding.
Additionally, in our table we use \npcc\ instead of ``R'' (resistant) and \p\ instead of ``V'' (vulnerable) or ``I'' (immune).
    \begin{observation}\label{obs:winner-model}
The 
complexities
of each of the standard
control problems
shown in Bartholdi, Tovey, and Trick~\shortcite{bar-tov-tri:j:control}
and Hemaspaandra, Hemaspaandra, and Rothe~\shortcite{hem-hem-rot:j:destructive-control} for the unique-winner model hold also for the nonunique-winner model.
\end{observation} }
In this section we will show that the
``M+,'' ``CF,'' and ``MF'' cases whose control type is classified as \p\ in Table~\ref{tbl:cm:specific},
are in the with-no-manipulators case in \p\ for each of our cooperative and competitive cases.\footnote{The reason we
have looked at only the P cases of control for these systems is 
    that due to our inheritance results, for the NP cases, getting a P
    result will be impossible.}

\begin{table}[h!]
\centering
\scriptsize %
 \def\arraystretch{1.5}
\begin{tabular}{|l|l|l|l|l|l|l|}
\hline
 & \multicolumn{2}{c|}{Plurality} & \multicolumn{2}{c|}{Condorcet} & \multicolumn{2}{c|}{Approval} \\ \cline{2-7}
Control by           & Constr.\ & Destr.\ & Constr.\ & Destr.\ &  Constr.\ & Destr.\ \\ \hline
Adding Candidates    & \npcc & \npcc & \p & \p & \p & \p \\ \hline
Deleting Candidates  & \npcc & \npcc & \p & \p & \p & \p \\ \hline
Adding Voters        & \p & \p & \npcc & \p & \npcc & \p \\ \hline
Deleting Voters      & \p & \p & \npcc & \p & \npcc & \p \\ \hline
\begin{tabular}[c]{@{}l}Partitioning\\ Candidates\end{tabular}& \begin{tabular}[c]{@{}l} TE: \npcc\\ TP: \npcc \end{tabular} & \begin{tabular}[c]{@{}l} TE: \npcc\\ TP: \npcc \end{tabular}  & \p  & \p &  \begin{tabular}[c]{@{}c} TE: \p \\ TP: \p\end{tabular} &  \begin{tabular}[c]{@{}c} TE: \p \\ TP: \p\end{tabular}  \\ \hline %
\begin{tabular}[c]{@{}l}Runoff Partitioning\\ Candidates\end{tabular}& \begin{tabular}[c]{@{}l} TE: \npcc\\ TP: \npcc \end{tabular} & \begin{tabular}[c]{@{}l} TE: \npcc\\ TP: \npcc \end{tabular}  & \p  & \p & \begin{tabular}[c]{@{}l} TE: \p \\ TP: \p\end{tabular} & \begin{tabular}[c]{@{}l} TE: \p \\ TP: \p\end{tabular}  \\ \hline %
\begin{tabular}[l]{@{}l}Partitioning \\Voters\end{tabular}& \begin{tabular}[c]{@{}l} TE: \p\\ TP: \npcc \end{tabular} & \begin{tabular}[c]{@{}l} TE: \p\\ TP: \npcc \end{tabular}  & \npcc  & \p & \begin{tabular}[c]{@{}l} TE: \npcc \\ TP: \npcc\end{tabular} & \begin{tabular}[c]{@{}l} TE: \p \\ TP: \p\end{tabular}  \\ \hline %
\end{tabular}
\caption{Summary of complexity of control for plurality, Condorcet, and approval~\cite{hem-hem-rot:j:destructive-control}.}
\label{tbl:cm:specific}
\end{table}

\begin{theorem}\label{thm:specific}
  Each problem contained in
  \begin{itemize}
  \item \mcontrol{
      \vtriple{approval}{Condorcet}{plurality}}{\vpaircd{}C \vpair{
        \vpairad\vpaircv }{ \vtriplestd\lhbox{-}\vpairtetp }},
  \item \scontrolcf{
      \vtriple{approval}{Condorcet}{plurality}}{\vpaircd{}C \vpair{
        \vpairad\vpaircv }{ \vtriplestd\lhbox{-}\vpairtetp }}, or
  \item \scontrolmf{
      \vtriple{approval}{Condorcet}{plurality}}{\vpaircd{}C \vpair{
        \vpairad\vpaircv }{ \vtriplestd\lhbox{-}\vpairtetp }},
  \end{itemize}
  whose corresponding control type is in $\p$ in Table~\ref{tbl:cm:specific}
    is in \p.
\end{theorem}

The proofs of many of these cases will utilize the polynomial-time algorithms
for the without-manipulators versions of the control cases. The well-known
polynomial-time results from
Bartholdi, Tovey, and Trick~\shortcite{bar-tov-tri:j:control} and
Hemaspaandra, Hemaspaandra, and Rothe~\shortcite{hem-hem-rot:j:destructive-control}
are both for the unique-winner model. Observation~\ref{obs:winner-model} states that each
of these control cases holds for the nonunique-winner model, and we will reference
this observation when referring to the polynomial-time algorithm for a given
nonmanipulator control case.

As an illustration, we present the proof of
\mbox{plurality-M+CCPV-TE$\,\,\in\p$} here. The proofs of the
remaining cases of Theorem~\ref{thm:specific} can be found
in Appendix~\ref{app:specific}.

\smallskip

\begin{proofs}\label{plurality-M+CCPV-TE}
  Note that it is not the case that the manipulators can always simply
  vote for $p$, no matter what the chair does.  For example, if the
  chair partitions the voters such that one of the subelections
  contains a voter voting $p > a > b$, and the other subelection
  contains 100 voters voting $a > b > p$, 101 voters voting $b > a >
  p$, and one manipulator, the manipulator should vote for $a$, so
  that $a$ and $b$ are tied in the second subelection and neither goes
  through to the second round.  Still, we will show that if a
  partition of the voters and a manipulation of the manipulators exist such
  that $p$ wins the election, then there exists a way for $p$ to win
  when all manipulators vote for $p$.  It follows that we can check if
  $p$ can be made a winner by first having all manipulators vote for
  $p$ and then running the polynomial-time algorithm for
  \scontrol{plurality}{CCPV-TE} from
  \cite{hem-hem-rot:j:destructive-control} (modified in the obvious way for the nonunique-winner case).

  So, suppose that a manipulation and a partition
  $(V_1,V_2)$ exist such that $p$ is a winner of the election.  Without loss
  of generality, suppose $p$ is the 
  unique winner of $(C,V_1)$.  Then
  $p$ is also the unique winner of $(C,V_1)$ if all manipulators in
  $V_1$ vote for $p$, so have all manipulators in $V_1$ vote for $p$. 
  Now consider $(C,V_2)$.
  As
  explained in the previous paragraph, simply changing the manipulators' votes to $p$
  could have bad effects.  Instead, we do the following.
  While manipulators remain in $V_2$ whose first-choice candidate is not $p$, choose one
  of them, $v$, let $a$ be $v$'s first-choice candidate, and do the following.
  \begin{enumerate}
    \item Change $v$'s vote from $a$ to $p$ and move $v$ to $V_1$.
    \item For each candidate $b \neq a$, move a current $V_2$ voter for $b$ (if any exists)
    from $V_2$ to $V_1$ and if it is a manipulator, change its vote to $p$.
  \end{enumerate}
  Since in each iteration of the above loop we add at least one vote
  for $p$ to $V_1$, $p$ will remain the unique winner of $(C,V_1)$.
  If after the loop $(C,V_2)$ does not have a unique winner or has $p$ as the unique winner
  it is immediate that $p$ wins the runoff.
  The only remaining case is that after the loop $(C,V_2)$ has a unique winner $c \neq p$.
  Note that in each iteration we keep the same set of winners in $(C,V_2)$ unless
  $V_2$ becomes empty in which case all candidates become winners in $(C,V_2)$.
  This implies that $c$ is the unique winner of $(C,V_2)$ before the loop and thus
  $c$ does not beat $p$ in the runoff before the loop.
  Since the only votes that are changed in the loop are manipulator votes changed to $p$,
  after the loop $p$ clearly is a winner of the runoff.\footnote{
  There was a slight problem in the argument used 
  in this paragraph in a previous 
  version~\cite{fit-hem-hem:c:control-manipulation,fit-hem-hem:tOutByConfWITHVERSION:control-manipulation}, which was fixed in
a later version~\cite{fit-hem-hem:c:control-manipulation-comsoc}.}
\end{proofs}

We now will seem to change directions, and will briefly study
``standard'' control problems, i.e., ones not in the presence of
manipulators. However, we do so in service of the goals of this paper.
The results we will obtain below will be crucially used to prove 
parts of 
Theorem~\ref{thm:specific}, though the proofs that do so 
are found not in the body of the paper but
in four proofs in~\ref{app:specific} that draw on the results below.

Below we state general results on election systems satisfying the Weak Axiom of Revealed Preferences (WARP) and
its corresponding unique version (unique-WARP).
An election system satisfies WARP if whenever a candidate is a winner among a set of candidates (under a
vote set $V$; as always, we assume that $V$ is masked down to the candidates at hand in the given election)
then that candidate is also a winner among every subset of those candidates that includes him or her
(under that same vote set $V$; as always, we assume
that $V$ is masked down to the candidates at hand in the given election).
Similarly, an election system satisfies unique-WARP if whenever a candidate is a unique winner
among a set of candidates
then that candidate is also a unique winner among every subset
of 
those candidates that includes him or %
her.\footnote{We here and 
in many other places write the somewhat strange, awkward phrase 
``a unique winner'' rather than the seemingly more natural phrase
``the unique winner.'' We do so to avoid giving the impression that there 
necessarily \emph{is} a unique winner---as opposed for example to 
perhaps having no winners or perhaps having multiple winners.

We mention in passing that WARP itself is very closely connected to 
immunity to destructive control by deleting
candidates (DCDC); in particular, they are the same.
To see this, we need to discuss a notion from the literature: immunity.  
An election system is said to be immune to 
destructive control by deleting
candidates if for every election instance $(C,V)$ and every candidate $c \in C$ it holds that: If $c$ is a winner 
in that election instance, then for every candidate set $C'$ satisfying $\{c\} \subseteq C'  \subseteq  C$ it holds that $c$ 
is a winner in the election with candidate set $C$ and vote set $V$ (masked down to the candidates in $C'$).
This notion, 
destructive control by deleting candidates, 
is due to the seminal control paper of Bartholdi, Tovey, and Trick~\cite{bar-tov-tri:j:control}, except their 
paper is in the unique-winner model and our paper is in the nonunique-winner model.
Yang~\cite{yan:c:control-bribery-majority-judgment} has observed that WARP implies, in the nonunique-winner model, immunity to 
destructive control 
by deleting candidates.  We here add the observation that the converse also holds, 
since the definitions of the two concepts are in fact the same.  Thus the following holds.
\begin{quote}
An election system \elec\ satisfies WARP if and only if \elec\ is immune to DCDC (destructive control 
by deleting candidates).
\end{quote}
Again, like all the results in this paper, the above if and only if statement is with respect to the nonunique-winner model.
We mention, for context, that in the unique-winner model (which is not 
the model we are using in this paper), the analogous result holds if one looks instead  
at unique-WARP, namely, we have the following result.
\begin{quote}
An election system \elec\ satisfies unique-WARP if and only if \elec\ is, in the unique-winner model, 
immune to DCDC (destructive control by deleting candidates).
\end{quote}
This result's ``only if'' direction is stated in~\cite{hem-hem-rot:j:destructive-control} and this result's 
``if'' direction clearly also holds, again as the definitions of the two notions in fact are the same.
Finally,
Yang~\cite{yan:c:control-bribery-majority-judgment}
(respectively, Hemaspaandra, Hemaspaandra, and Rothe~\cite{hem-hem-rot:j:destructive-control})
states that in the nonunique-winner model (respectively, unique-winner model), that WARP (respectively, unique-WARP)
implies immunity to constructive control by adding candidates.  We observe that the converse directions
for each of those claims hold, for the same reasons as mentioned above for the DCDC cases, thus yielding two
additional if and only if results.}
It is easy to see that approval and
Condorcet elections satisfy both WARP and unique-WARP~\cite{hem-hem-rot:j:destructive-control}.

Though 
as mentioned above 
these results are 
rather crucially 
used as tools within our proofs 
about control in the presence of manipulators, we feel they are of
interest in their own right.
Let us take as an example the coming Theorem~\ref{thm:warp-one-part}, which
loosely put says that for every election system satisfying 
unique-WARP, and
for each instance of 
CCRPC-TE,
it
holds that 
the partition whose parts are
``all candidates other than $p$'' and ``$p$'' will cause $p$ to win
if and only if the chair has \emph{any} partition choice that will
cause $p$ to win.

The result is interesting because it is directly attacking what
is the heart of the complexity of partition problems: 
combinatorial explosion, i.e., 
the fact that 
there are an enormous number of partitions and the 
chair must
determine whether any one of them makes $p$ a winner.  This is
precisely why such problems so often turn out to be NP-hard.  However,
Theorem~\ref{thm:warp-one-part} says that for systems obeying the
unique-WARP axiom, that potential complexity is completely
side-stepped: There is a single partition that is the only one that
needs to be examined.  This immediately shows that the control type 
is of polynomial-time complexity for systems satisfying unique-WARP\@.  

Viewed more broadly, by linking the complexity of control to
social-choice properties, this part of our work is trying to take a
step away from analyzing systems one at a time, and is trying to more
generally determine what it is that can yield computational
simplicity.  
Work having that goal is 
most typically done 
by studying the class of so-called scoring systems, each
of which is defined by a so-called scoring vector, and finding some
simple property of the scoring vector that determines the
complexity of various manipulative attacks on elections.  To give
as an example just one family of such results, we mention the line doing
this regarding manipulation of elections in the general case and in
the so-called single-peaked
case~\cite{con-lan-san:j:when-hard-to-manipulate,hem-hem:j:dichotomy-scoring,pro-ros:j:juntas,fal-hem-hem-rot:j:single-peaked-control,bra-bri-hem-hem:j:single-peaked-bribery}.
However, that work focuses on the direct definitions of the election
systems, and our work 
in contrast is focusing on
how possession of an axiomatic property can itself 
force simplicity.

Let us now turn to our results of this type.

\begin{theorem}\label{thm:warp-one-part}
For every election system \elec\ satisfying unique-WARP,
and for each instance of the CCRPC-TE problem, 
it holds that control is possible if and only if 
the preferred candidate $p$ is an overall winner using the partition $(C-\{p\},\{p\})$.
\end{theorem}

\begin{proofs}
Given an election system satisfying unique-WARP, an election $(C,V)$,
and a candidate $p \in C$, we do the following.

If $p$ is an overall winner using partition $(C-\{p\},\{p\})$ then clearly
control is possible.

Conversely, %
if $p$ is \emph{not} an
overall winner using partition $(C-\{p\},\{p\})$ then we will show that control
is not possible.
There are two cases.  
\begin{enumerate}
\item If 
under our set of votes (masked down 
to the candidates in the election at hand in each case, of course)
$p$ does not win in the election where $p$ is the sole candidate,
then by unique-WARP $p$ will not be a unique winner in any subelection %
it is part of,
and so can never survive the first round, and so can never become 
an overall winner.
\item On the other hand, if 
under our set of votes (masked down 
to the candidates in the election at hand in each case, of course)
$p$ wins in the election where $p$ is the sole candidate, then 
$p$ in the partition 
$(C-\{p\},\{p\})$ 
clearly will survive the first round.  

Since we are in the TE model,
either zero or one candidates will survive the 
$C-\{p\}$ first-round subelection.  

But if zero survive, then the 
second-round election involves just $p$, who we already, in 
our current case, have assumed wins 
under the votes masked down to it, so it will in fact be an
overall winner (in fact, it will be the only overall winner).

On the other hand, if one candidate, call it $r$, 
survives the 
$C-\{p\}$ first-round subelection, 
note that 
since we assumed that $p$ is not an overall winner, it must be 
the case that in the election between $r$ and $p$ (with the votes 
as always masked down to the candidates in the election), 
$p$ is not a winner.  So, can there be any partition, $(C-A, A)$, under the 
given votes, that will ensure that $p$ is an overall winner?
W.l.o.g., assume $p \in A$.
If $r \in A$, then $p$ cannot move forward, since to do that 
(as we are in the TE model) $p$ would have to be a unique 
winner within $A$, and since $\{p,r\}\subseteq A$, by 
unique-WARP
it would have been impossible for $p$ to fail to beat $r$ in the second-round 
election under 
partition $(C-\{p\},\{p\})$ in our original setting, 
yet that is precisely what happened in our current
case's assumptions.  On the other hand, if $r \not\in A$, then 
given that $C - A \subseteq C - \{p\}$, by unique-WARP we have that 
$r$ wins the subelection $(C-A,V)$, %
and so faces $p$ in the
runoff, %
and we already know that in that case $p$ will not be a winner 
of that contest.
\end{enumerate}
By the above case analysis, we have shown that 
control is not possible, thus completing this second 
direction of the proof.~\end{proofs}

\begin{corollary}\label{cor:warp-one-part-complexity}
For every election system \elec\ that satisfies unique-WARP and has a p-time winner problem,
$\scontrol{\elec}{CCRPC-TE}$ is in \p.
\end{corollary}

Theorem~\ref{thm:warp-one-part} does not hold for CCPC-TE\@. For example, in the election system where all
candidates are winners if there are at least two candidates, and no candidates win if there is at most one candidate
(note that this system vacuously satisfies unique-WARP), an election with candidates $\{a,b\}$ has no winners
using partition $(\{a\},\{p\})$, but all candidates win using partition $(\emptyset,\{a,p\})$. 

Nonetheless, we have proven an
analogue of Theorem~\ref{thm:warp-one-part} for the 
CCPC-TE case.  Our analogue, however, 
applies to election systems that satisfy 
both WARP and unique-WARP\@.\footnote{Is it going unnaturally far to 
study systems that satisfy both WARP and unique-WARP\@?  We do 
not think so.  Indeed, to put our use of two 
properties in context, we mention that even combined they 
are a weaker assumption about the election system than is even 
a certain different version of WARP that is sometimes used.
The version of 
WARP that we are using here is precisely
that found for example in Baumeister and Rothe's 
survey of preference 
aggregation~\cite{bau-rot:b:preference-aggregation-by-voting}.  
This version focuses on the individual
candidate and what happens when other candidates are removed, 
namely, that winning does not turn into not winning for any 
unremoved candidate.  The other version, and to avoid 
confusion let us refer to it as WARP$'$, focuses on whether 
when one removes candidates the winner set is always 
\emph{exactly} the previous winner set intersected 
with the remaining set of candidates.  WARP$'$ clearly 
implies both WARP and unique-WARP\@.  And so 
Theorem~\ref{thm:warp-two-part} would certainly 
remain true if in it one were to 
replace the phrase 
``both WARP and unique-WARP'' with simply ``WARP$'$''.}

\begin{theorem}\label{thm:warp-two-part}
For every election system satisfying both WARP and unique-WARP, 
and for each instance of the CCPC-TE problem, 
it holds that control is possible if and only if 
the preferred candidate $p$ is an overall winner using the partition $(C-\{p\},\{p\})$.
\end{theorem}

\begin{proofs}
Given an election system satisfying both WARP and unique-WARP, an election $(C,V)$, and a candidate $p \in C$,
we do the following.

If $p$ is an overall winner using partition $(C-\{p\},\{p\})$ then clearly control is possible.

Conversely, if $p$ is \emph{not} an overall winner using partition $(C-\{p\},\{p\})$ then we will show that control is not
possible. There are two cases.

\begin{enumerate}

\item If under our set of votes (masked down to the candidates in the election at hand in each case, of course) $p$
does not win in the election where $p$ is the sole candidate, then by WARP $p$ will not be a winner in any larger
subelection that contains him or her, and so can never be an overall winner.

\item If under our set of votes (masked down to the candidates in the election at hand in each case, of course) $p$ wins
the election where $p$ is the sole candidate then since $p$ is not the overall winner using partition
$(C-\{p\},\{p\})$ and we are in the TE model, there exists a candidate $r \in C-\{p\}$ such that $r$ is the unique winner
of the subelection $(C-\{p\},V)$ and $p$ does not win the runoff election $(\{p,r\},V)$.
Since the given election system satisfies unique-WARP and $r$ is the unique winner of $(C-\{p\},V)$,
$r$ will be the unique winner of every subelection that does not involve $p$. And since the given election satisfies
WARP and $p$ does not win $(\{p,r\},V)$, $p$ is not a winner in any subelection involving $r$. Notice that $p$ participates
in the runoff only if $r$ also participates in the runoff. So it is clear to see that control is not possible.
\end{enumerate}
\end{proofs}

\begin{corollary}\label{cor:warp-two-part-complexity}
For every election system \elec\ that satisfies both WARP and unique-WARP and has a p-time winner problem,
$\scontrol{\elec}{CCPC-TE}$ is in \p.
\end{corollary}

\begin{corollary}\label{cor:warp-part-models}
For every election system \elec\ satisfying both WARP and unique-WARP,
$\scontrol{\elec}{CCPC-TE} = \scontrol{\elec}{CCRPC-TE}$.
\end{corollary}

\subsubsection{Weighted Voters}\label{ss:weighted}

We now give results for veto and Borda, including, for the latter, an
interesting increase in complexity.

In weighted
elections every voter has a positive integer weight, and a voter with
weight $w$ counts as $w$ voters.
In weighted voter control cases, the addition/deletion limit still
pertains to the number of voters that can be added or deleted.
Consider the case of 3-candidate weighted veto elections.  The known
results on this are that constructive coalitional manipulation is
NP-complete~\cite{con-lan-san:j:when-hard-to-manipulate},
destructive coalitional manipulation is in P~\cite{con-lan-san:j:when-hard-to-manipulate},
and CCAV 
and CCDV are both in $\p$~\cite{fal-hem-hem:j:weighted-control}.
The following result, whose second part
may be surprising, shows that for this system
CC\vpairad{}V-\vpair{CF}{MF} are all in $\p$---not $\np$-complete.
\begin{theorem}\label{t:veto-surprise}For 3-candidate weighted veto
  elections, the following hold.
  \begin{enumerate}
  \item {\rm M+CC\vpairad{}V\/} are both $\np$-complete.
  \item {\rm CC\vpairad{}V-\vpair{CF}{MF}\/} are each in $\p$.
  \end{enumerate}
\end{theorem}

\begin{proofs}
  The first case follows directly from the fact that constructive
  manipulation is
  \np-complete~\cite{con-lan-san:j:when-hard-to-manipulate} and the
  inheritance observations from Section~\ref{ss:inheritance} (as the
  relevant result there holds even for the weighted case).

  For the competitive cases, note that the only action that makes
  sense for the manipulators is to veto $p$.  This holds regardless of
  whether the manipulators or the chair goes first.  So, we let the
  manipulators veto $p$ and then run the polynomial-time algorithm for
  CCAV and CCDV from~\shortcite{fal-hem-hem:j:weighted-control}.%
\end{proofs}

3-candidate weighted Borda elections show a true increase in
complexity.
The known
results for this system are that constructive coalitional manipulation is
NP-complete~\cite{con-lan-san:j:when-hard-to-manipulate},
destructive coalitional manipulation is in P~\cite{con-lan-san:j:when-hard-to-manipulate},
and CCAV 
and CCDV are both \np-complete~\cite{fal-hem-hem:j:weighted-control} and thus
all these problems are in \np.
Yet we show that CCAV-MF is
coNP-hard, and so cannot be in NP unless the polynomial hierarchy
collapses to $\np \cap \conp$.

\begin{theorem}\label{t:borda-surprise}
  For 3-candidate weighted Borda elections, the following hold.
  \begin{enumerate}
  \item {\rm M+CC\vpairad{}V\/} are both $\np$-complete.
  \item {\rm CC\vpair{AV\hbox{-}CF}{DV\hbox{-}\vpaircfmf}} are each
    $\np$-hard.
  \item {\rm CCAV-MF} is $\np$-hard and $\conp$-hard.
  \item {\rm CC\vpairad{}V\hbox{-}CF\/} is $\np$-complete.
  \end{enumerate}
\end{theorem}

\begin{proofs}
  The first case follows directly from the fact that manipulation is
  \np-complete~\cite{con-lan-san:j:when-hard-to-manipulate} and the
  inheritance observations from Section~\ref{ss:inheritance}.

  The remaining NP-hardness results follow from the \np-completeness
  of CCAV and CCDV and the inheritance observations from
  Section~\ref{ss:inheritance}.

  To show that CCAV-CF is in \np, guess a set of voters to add, and
  then check that the manipulators can't make $p$ not win. We do this
  by setting all manipulators to $a > b > p$, checking that $p$ is a
  winner, and then setting all manipulators to $b > a > p$, and
  checking that $p$ is a winner.  A similar argument shows that
  CCDV-CF is in \np.

  It remains to show that CCAV-MF is \conp-hard, i.e., that the
  complement of CCAV-MF is \np-hard.  We will reduce from Partition.
  Given a nonempty sequence of positive integers $k_1, \ldots, k_t$
  that sums to $2K$, we will construct an election such that there is
  a partition (i.e., a subsequence of $k_1, \ldots, k_t$ that sums to
  $K$) if and only if the manipulators can vote in such a way that the
  chair won't be able to make $p$ a winner.

  We construct the following election: We have manipulators with
  weights $k_1, \ldots, k_t$.  The manipulators are registered voters.
  We have two unregistered voters,
  both with weight $3K - 1$.  One of these voters votes $p > a > b$
  and one votes $p > b > a$.  We have addition limit one, i.e., the
  chair can add at most one voter.

  If there is a partition, then the manipulators vote so that a total
  of $K$ vote weight casts the vote $a > b > p$ and a total of $K$
  vote weight casts the vote $b > a > p$.  So, the scores of $p$, $a$,
  and $b$ are $0$, $3K$, and $3K$.  There is no way for the chair to
  make $p$ a winner by adding at most one voter.  If the chair adds
  the weight $3K-1$ voter voting $p > a > b$, the score of $p$ is
  $6K-2$ and the score of $a$ is $3K + (3K-1) = 6K -1$ and so $p$ is
  not a winner.  Adding the other voter gives a score of $6K-2$ for
  $p$ and a score of $6K-1$ for $b$ and again $p$ is not a winner.

  Now consider the case that there is no partition.  Look at the
  scores of the candidates after the manipulators have voted.  Without
  loss of generality, assume that $\score{a} \leq \score{b}$. Then
  $\score{a} \leq 3K - 1$ (since there is no partition) and $\score{b}
  \leq 4K$.  Now the chair adds the weight $3K-1$ voter voting $p > a
  > b$.  After adding that voter, $p$'s score is $6K - 2$, $a$'s score
  is at most $(3K -1) + (3K-1)$ and $b$'s score is at most $4K$.  It
  follows that $p$ is a winner.%
\end{proofs}

\section{Conclusions and Open Directions}
We have established general inheritance results and complexity upper
bounds for control in the presence of manipulators, for both
cooperative and competitive settings.  We for the upper bounds
provided matching lower bounds, but also showed that for many natural
systems the complexity is far lower than the general upper bounds.

Many open directions remain.  For example, regarding 3-candidate
weighted Borda elections, we have shown that CCAV-MF is NP-hard and
coNP-hard, and although our upper-bound theorem is not explicitly
about weighted cases, clearly this problem, for exactly the same
reason as in our upper-bound theorem, is in $\conpnp$.  But precisely
where within that range does it fall?  Also, what happens for
real-world election systems that themselves are complex to manipulate
and/or control, such as Llull, Copeland, fallback, sincere-preference
approval, and Schulze elections?  Do some of these systems themselves
provide natural systems that might for our competitive cases be
complete for some of the high complexity classes given in
Table~\ref{t:upper-general}?

\bigskip

\noindent
\subsection*{Acknowledgments} 
We are grateful to the anonymous COMSOC and IJCAI
referees 
for helpful comments and suggestions.

\bibliographystyle{alpha} %
\newcommand{\etalchar}[1]{$^{#1}$}

\appendix

\section{Deferred Proofs from Section~\ref{sec:upper-general-lower}}
\label{app:upperbound}

\begin{theorem}\label{CCAC-MF:complete}
There exists an election system, \elec, with a p-time winner problem,
such that \escontrolmf{CCAC} is \pitwo-complete.\end{theorem}

\begin{proofs}
Let \elec\ be as defined in the proof of Theorem~\ref{CCAC-CF:complete}.
Then \elec\ has a p-time winner problem and by
Theorem~\ref{t:upper-bounds} we know that \escontrolmf{CCAC}
is in \pitwo. So what is left is to show that \escontrolmf{CCAC}
is \pitwo-hard.

Let $(\forall x_{\ell+1}, \ldots , x_{2\ell})(\exists x_{1},\ldots, x_{\ell})
[\psi(x_1,\ldots,x_{2\ell})]$ be an instance of $\qbfpi{2}$.
Our instance of \escontrolmf{CCAC} is exactly the instance of
\escontrolcf{CCAC} from the proof of Theorem~\ref{CCAC-CF:complete}.
The same argument as in that proof shows that 
$(\forall x_{\ell+1}, \ldots, x_{2\ell})(\exists x_{1},\ldots, x_{\ell})
[\psi(x_1,\ldots,x_{2\ell})] \in \qbfpi{2}$ if and only if the
chair can always ensure that $p$ becomes a winner.~\end{proofs}

\begin{theorem}\label{CCDC-CF:complete}
There exists an election system, \elec, with a p-time winner problem,
such that \escontrolcf{CCDC} is \sigmatwo-complete.\end{theorem}

\begin{proofs}
Let \elec\ be defined as in the proof of Theorem~\ref{CCAC-CF:complete}.
Then \elec\ has a p-time winner problem and by
Theorem~\ref{t:upper-bounds} we know that \escontrolcf{CCDC}
is in \sigmatwo. So what is left is to show that \escontrolcf{CCDC}
is \sigmatwo-hard.

Let $(\exists x_1, \ldots, x_\ell)(\forall x_{\ell+1},\ldots, x_{2\ell})
[\psi(x_1,\ldots,x_{2\ell})]$ be an instance of $\qbfsigma{2}$.
Our instance of \escontrolmf{CCDC} is the instance of
\escontrolcf{CCAC} from the proof of Theorem~\ref{CCAC-CF:complete},
except that we let the candidate set $C$ consist of all $2\ell +1$ candidates.
The same argument as in the proof of Theorem~\ref{CCAC-CF:complete}
shows that 
$(\exists x_1, \ldots, x_\ell)(\forall x_{\ell+1},\ldots, x_{2\ell})
[\psi(x_1,\ldots,x_{2\ell})] \in \qbfsigma{2}$ if and only if
the chair can ensure that $p$ always becomes a winner by
deleting candidates.~\end{proofs}

\begin{theorem}\label{CCDC-MF:complete}
There exists an election system, \elec, with a p-time winner problem,
such that \escontrolmf{CCDC} is \pitwo-complete.\end{theorem}

\begin{proofs}
Let \elec\ be defined as in the proof of Theorem~\ref{CCAC-CF:complete}.
Then \elec\ has a p-time winner problem and by
Theorem~\ref{t:upper-bounds} we know that \escontrolmf{CCDC}
is in \pitwo. So what is left is to show that \escontrolmf{CCDC}
is \pitwo-hard.

Let $(\forall x_{\ell+1}, \ldots, x_{2\ell})(\exists x_{1},\ldots, x_{\ell})
[\psi(x_1,\ldots,x_{2\ell})]$ be an instance of $\qbfpi{2}$.
Our instance of \escontrolmf{CCDC} is exactly the instance of
\escontrolcf{CCDC} from the proof of Theorem~\ref{CCDC-CF:complete}.
The same argument as in that proof of Theorem~\ref{CCAC-CF:complete}
shows that 
$(\forall x_{\ell+1}, \ldots, x_{2\ell})(\exists x_{1},\ldots, x_{\ell})
[\psi(x_1,\ldots,x_{2\ell})] \in \qbfpi{2}$ if and only if the
chair can ensure that $p$ always becomes a winner.~\end{proofs}

\begin{theorem}\label{DC:complete}
There exists an election system, \elecprime, with a p-time winner problem,
such that
\scontrolcf{\elecprime}{DCAC} is \sigmatwo-complete,
\scontrolmf{\elecprime}{DCAC} is \pitwo-complete,
\scontrolcf{\elecprime}{DCDC} is \sigmatwo-complete, and
\scontrolmf{\elecprime}{DCDC} is \pitwo-complete.
\end{theorem}

\begin{proofs}
Let \elecprime\ be defined as $\elec$ in Theorem~\ref{CCAC-CF:complete} except
replace %
``everyone loses'' with ``everyone wins'' and
``everyone wins'' with ``everyone loses.''

Note that for every election $(C,V)$ and every candidate $p \in C$,
$p$ is an \elec\ winner of $(C,V)$ if and only if $p$ is not an
\elecprime\ winner of $(C,V)$. This immediately implies that
\scontrolcf{\elecprime}{DCAC} = \scontrolcf{\elec}{CCAC},
\scontrolmf{\elecprime}{DCAC} = \scontrolmf{\elec}{CCAC},
\scontrolcf{\elecprime}{CCDC} = \scontrolcf{\elec}{CCDC}, and
\scontrolmf{\elecprime}{CCDC} = \scontrolmf{\elec}{CCDC}.
The result follows from
Theorems~\ref{CCAC-CF:complete},~\ref{CCAC-MF:complete},~\ref{CCDC-CF:complete}, and~\ref{CCDC-MF:complete}.~\end{proofs}

\begin{theorem}\label{CCAV-CF:complete}
There exists an election system, \elec, with a p-time winner problem,
such that \escontrolcf{CCAV} is \sigmatwo-complete.\end{theorem}

\begin{proofs}
Let \elec\ be defined in the following way. Given an election $(C,V)$,
if $\|C\| \ge 3$ and the
candidates in $C$ listed in increasing lexicographic order %
are $c_0, c_1, \dots, c_{\ell+1}$, %
candidate $c_0$ encodes a boolean formula
$\psi(x_1, \ldots, x_{2\ell})$, $\|V\| = 2\ell+1$, and for each
$i, 1 \le i \le \ell$ there are at least two voters with the same vote
who rank $c_i$ first, then do the following.
For each $i$, $1 \le i \le \ell$, set $x_i$ to true if two voters with $c_i$
first both state $c_{\ell + 1} > c_0$ and otherwise set $x_i$ to false.
Let $\widehat{v}$ be the unique vote that occurs three times
or only once in $V$.
For each $i$, $1 \le i \le \ell$, set $x_{\ell+i}$ to true if $\widehat{v}$
states $c_{i} > c_0$, else set $x_{\ell+i}$ to false.
If this is a satisfying assignment for $\psi$ then everyone wins.
In all other cases everyone loses.
That completes the specification of \elec.

Clearly \elec\ has a p-time winner problem, and by Theorem~\ref{t:upper-bounds}
we know that \escontrolcf{CCAV} is in \sigmatwo. So what is left is to show
that \escontrolcf{CCAV} is \sigmatwo-hard.

Let $(\exists x_{1}, \ldots, x_{\ell})(\forall x_{\ell + 1},\ldots, x_{2\ell})
[\psi(x_1,\ldots,x_{2\ell})]$ be an instance of \qbfsigma{2}.
We construct an instance of \escontrolcf{CCAV} in the following way.
Let the candidate set $C$ consist of $p$ encoding $\psi$ and $\ell+1$ candidates
all lexicographically larger than $p$. So, the candidates in $C$ can be listed
in increasing lexicographic order as $p, c_1, \dots, c_{\ell + 1}$.
Let the collection of registered voters $V$ consist of zero nonmanipulators
and one manipulator.
Let the collection of unregistered voters, all nonmanipulators,
consist of $2\ell$ pairs where
for each $i, 1 \le i \le \ell$, there are two voters
$v_i$ and $v'_i$ with the same vote $c_i > c_{\ell + 1} > p > \cdots$
and two voters $u_i$ and $u'_i$ with the same vote $c_i > p > c_{\ell + 1} > \cdots$.
Let the add limit $k = 4\ell$
and let the preferred candidate of the
chair be $p \in C$.

If $(\exists x_{1}, \ldots, x_{\ell})(\forall x_{\ell+1},\ldots, x_{2\ell})
[\psi(x_1,\ldots,x_{2\ell})] \in \qbfsigma{2}$, fix an assignment to
$x_1, \dots, x_{\ell}$ such that
$(\forall x_{\ell+1},\dots,x_{2\ell})[\psi(x_1,\dots,x_{2\ell})]$ is true.
For each $i, 1 \le i \le \ell$, if $x_i$ is true in the assignment the chair
adds $v_i$ and $v'_i$ and if $x_i$ is false the chair adds
$u_i$ and $u'_i$.
Note that the vote of the manipulator will be the unique vote 
$\widehat{v}$ that occurs three times (if the manipulator votes 
the same as one of the paired voters) or only once.
And no matter what assignment to
$x_{\ell + 1}, \ldots, x_{2\ell}$ is induced by $\widehat{v}$,
formula $\psi$ is satisfied and so $p$ will win.

Conversely, if the chair makes $p$ a winner then the chair adds exactly $\ell$
voter pairs whose $\ell$ different votes give an assignment to
$x_1, \dots, x_\ell$ such that
$(\forall x_{\ell+1}, \dots, x_{2\ell})[\psi(x_1,\dots,x_{2\ell})]$
is true.~\end{proofs}

\begin{theorem}\label{CCAV-MF:complete}
There exists an election system, \elec, with a p-time winner problem,
such that \escontrolmf{CCAV} is \pitwo-complete.\end{theorem}

\begin{proofs}
Let \elec\ be as defined in the proof of Theorem~\ref{CCAV-CF:complete}.
Then \elec\ has a p-time winner problem and by
Theorem~\ref{t:upper-bounds} we know that \escontrolmf{CCAV}
is in \pitwo. So what is left is to show that \escontrolmf{CCAV}
is \pitwo-hard.

Let $(\forall x_{\ell+1}, \ldots , x_{2\ell})(\exists x_{1},\ldots, x_{\ell})
[\psi(x_1,\ldots,x_{2\ell})]$ be an instance of $\qbfpi{2}$.
Our instance of \escontrolmf{CCAV} is exactly the instance of
\escontrolcf{CCAV} from the proof of Theorem~\ref{CCAV-CF:complete}.
The same argument as in that proof shows that 
$(\forall x_{\ell+1}, \ldots, x_{2\ell})(\exists x_{1},\ldots, x_{\ell})
[\psi(x_1,\ldots,x_{2\ell})] \in \qbfpi{2}$ if and only if the
chair can always ensure that $p$ becomes a winner.~\end{proofs}

\begin{theorem}\label{DC:avoter:complete}
There exists an election system, \elecprime, with a p-time winner problem,
such that
\scontrolcf{\elecprime}{DCAV} is \sigmatwo-complete and
\scontrolmf{\elecprime}{DCAV} is \pitwo-complete.
\end{theorem}

\begin{proofs}
Let \elecprime\ be defined as $\elec$ in Theorem~\ref{CCAV-CF:complete} except
replace %
``everyone loses'' with ``everyone wins'' and
``everyone wins'' with ``everyone loses.''

Note that for every election $(C,V)$ and every candidate $p \in C$,
$p$ is an \elec\ winner of $(C,V)$ if and only if $p$ is not an
\elecprime\ winner of $(C,V)$. This immediately implies that
\scontrolcf{\elecprime}{DCAV} = \scontrolcf{\elec}{CCAV} and
\scontrolmf{\elecprime}{DCAV} = \scontrolmf{\elec}{CCAV}.
The result follows from
Theorems~\ref{CCAV-CF:complete}
and~\ref{CCAV-MF:complete}.~\end{proofs}

Unlike in the candidate cases, we can not use the same construction 
to show that the deleting voter cases are also hard, because
the chair can delete the manipulator. In fact, 
we will show that for every election system \elec\ with a p-time winner problem,
\escontrolcf{CCDV}
and
\escontrolcf{DCDV}
 are in coDP (and so are not \sigmatwo-complete unless the 
polynomial hierarchy collapses).
DP is the class of languages that are the difference of two NP 
languages~\cite{pap-yan:j:dp}. 

\begin{theorem}\label{CDV-CF:coDP}
For every election system \elec\ with a p-time winner problem,
\escontrolcf{CCDV} and \escontrolcf{DCDV}
are in {\rm coDP}.\end{theorem}

\begin{proofs}
It is easy to see that it is always at least as good for the chair to delete
a manipulator as it is to delete a nonmanipulator (though note
that because the election system can be anything, deleting as many
manipulators as possible may not be best; for example, if we want to make
$p$ a winner and our election systems has all candidates as
winners if there are four voters and no winners if there are fewer voters,
we do not want to delete manipulators if there are four voters).
So we have that
$p$ can be made a winner (not a winner)
by deleting at most $k$ voters if and only if 
there exists a $k' \leq k$ such that (letting $m$
be the number of manipulators):
\begin{enumerate}
\item $k' \leq m$ and after deleting $k'$ manipulators the remaining 
$m-k'$ manipulators can not preclude $p$ from winning (not winning), or
\item $k' > m$ and after deleting all manipulators the chair can make
$p$ win (not win) by deleting at most $k' - m$ voters.
\end{enumerate}
We can check 
if there exists a $k'$ such that we are in case 1 in coNP and we can
check if there exists  
a $k'$ such that we are in case 2 in NP, and so we can write our
languages as the union of a coNP set and an NP set.
\end{proofs}

We now show that the coDP bounds from Theorem~\ref{CDV-CF:coDP} are tight.

\begin{theorem}\label{CCDV-CF:complete}
There exists an election system, \elec, with a p-time winner problem,
such that \escontrolcf{CCDV} is {\rm coDP}-complete.\end{theorem}

\begin{proofs}
We reduce from the coDP-complete
problem $\{\pair{\phi,\psi} \ | \ \phi \in \sat \mbox{ or }
\psi \not \in \sat\}$, which is the complement of the
standard DP-complete problem SAT-UNSAT~\cite{pap-yan:j:dp}.
Without loss of generality, we assume that $\phi$ and $\psi$ have the same 
number of variables.

Let \elec\ be defined in the following way. Given an election $(C,V)$,
if $\|C\| \ge 3$ and the
candidates in $C$ listed in increasing lexicographic order %
are $c_0, c_1, \dots, c_{\ell+1}$, %
and 
candidate $c_0$ encodes the pair of boolean formulas
$\pair{\phi(x_1, \ldots, x_\ell), \psi(x_{\ell+1}, \ldots, x_{2\ell})}$, then:
\begin{enumerate}
\item If $\|V\| = \ell$ and for each
$i, 1 \le i \le \ell$, there is a voter
who ranks $c_i$ first, we do the following.
For each $i$, $1 \le i \le \ell$, set $x_i$ to true if the voter with $c_i$
first states $c_{\ell + 1} > c_0$ and otherwise set $x_i$ to false.
If this is a satisfying assignment for $\phi$, then everyone wins.
\item If $\|V\| = 2\ell + 1$, then if there are no voters that rank
$c_{\ell+1}$ first, then everyone wins.
Otherwise, if there is exactly one voter $\widehat{v}$ that 
ranks $c_{\ell + 1}$ first then
for each $i, 1 \le i \le \ell$, set $x_{\ell+i}$ to true if $\widehat{v}$
states $c_{i} > c_0$, else set $x_{\ell+i}$ to false.
If this is not a satisfying assignment for $\psi$, then everyone wins.
\end{enumerate}
In all other cases everyone loses.
That completes the specification of \elec.

Clearly \elec\ has a p-time winner problem, and by Theorem~\ref{CDV-CF:coDP}
we know that \escontrolcf{CCDV}
is in {\rm coDP}. So what is left is to show that \escontrolmf{CCDV}
is {\rm coDP}-hard.

Let $\pair{\phi(x_1, \ldots, x_\ell), \psi(x_{\ell+1},\ldots,x_{2\ell})}$ be a
pair of boolean formulas.
We construct an instance of \escontrolcf{CCDV} in the following way.
Let the candidate set $C$ consist of $p$ encoding $\pair{\phi,\psi}$
and $\ell+1$ candidates
all lexicographically larger than $p$. So, the candidates in $C$ can be listed
in increasing lexicographic order as $p, c_1, \dots, c_{\ell + 1}$.
Let the collection of voters $V$ consist of one manipulator
and $2\ell$ nonmanipulators where
for each $i, 1 \le i \le \ell$, there is a voter
$v_i$ who votes $c_i > c_{\ell + 1} > p > \cdots$
and a voter $u_i$ who votes $c_i > p > c_{\ell + 1} > \cdots$.
Let the delete limit $k = 2\ell+1$
(any limit $\geq \ell+1$ will do)
and let the preferred candidate of the
chair be $p \in C$.
We need to show that 
($\phi \in \sat \mbox{ or } \psi \not \in \sat$) if and only if control
can be asserted.

Suppose $\phi \in \sat$. Fix an assignment to $x_1, \ldots, x_\ell$ 
that satisfies $\phi$. The chair deletes $\ell + 1$ voters.
The only voters that are not deleted are for each $i, 1 \leq i \leq \ell$,
$v_i$ if $x_i$ is true in the
assignment and $u_i$ if $x_i$ is false in the assignment.
This leaves $\ell$ voters that encode a satisfying assignment for $\phi$ and
so everyone wins.
Next suppose that $\psi \not \in \sat$. Then we keep all voters. 
Since there does not exist a satisfying assignment for $\psi$, 
everyone wins.

For the converse, to have $p$ win, we either have that $\|V\| = \ell$, in which
case $\phi$ is satisfiable, or $\|V\| = 2\ell + 1$. In the latter case,
if $\psi \in \sat$ the manipulator could induce a satisfying 
assignment for $\psi$, but then $p$ is not a winner. It follows that
$\psi \not \in \sat$.~\end{proofs}

\begin{theorem}\label{DCDV-CF:complete}
There exists an election system, \elecprime, with a p-time winner problem,
such that \escontrolcf{DCDV} is {\rm coDP}-complete.\end{theorem}

\begin{proofs}
Let \elecprime\ be defined as $\elec$ in Theorem~\ref{CCDV-CF:complete} except
replace %
``everyone loses'' with ``everyone wins'' and
``everyone wins'' with ``everyone loses.''

Note that for every election $(C,V)$ and every candidate $p \in C$,
$p$ is an \elec\ winner of $(C,V)$ if and only if $p$ is not an
\elecprime\ winner of $(C,V)$. This immediately implies that
\scontrolcf{\elecprime}{DCDV} = \scontrolcf{\elec}{CCDV}.
The result follows from
Theorem~\ref{CCDV-CF:complete}.~\end{proofs}

For the CCDV-MF case, we modify the construction from 
Theorem~\ref{CCAV-MF:complete} to basically ensure that
the manipulator will not be deleted, while still making
sure that $p$ can always be made a winner for positive 
instances of $\qbfpi{2}$.

\begin{theorem}\label{CCDV-MF:complete}
There exists an election system, \elec, with a p-time winner problem,
such that \escontrolmf{CCDV} is \pitwo-complete.\end{theorem}

\begin{proofs}
Let \elec\ be defined in the following way. Given an election $(C,V)$,
if $\|C\| \ge 3$ and the
candidates in $C$ listed in increasing lexicographic order %
are $c_0, c_1, \dots, c_{\ell+1}$, %
candidate $c_0$ encodes a boolean formula
$\psi(x_1, \ldots, x_{2\ell})$, $\|V\| = \ell+1$, and for each
$i, 1 \le i \le \ell$, there is a voter
who ranks $c_i$ first, then do the following.
For each $i$, $1 \le i \le \ell$, set $x_i$ to true if some voter with $c_i$
first states $c_{\ell + 1} > c_0$ and otherwise set $x_i$ to false.
If there is a voter $\widehat{v}$ that ranks $c_{\ell + 1}$ first 
(note that there exists at most one such voter) then
for each $i, 1 \le i \le \ell$, set $x_{\ell+i}$ to true if $\widehat{v}$
states $c_{i} > c_0$, else set $x_{\ell+i}$ to false.
If this is a satisfying assignment for $\psi$ then everyone wins.
If there is a voter that ranks $c_0$ first then everyone wins.
If there are two voters that rank $c_i$ first for some $i, 1 \leq i \leq \ell$,
and these voters 
agree on whether or not $c_{\ell + 1} > c_0$ then everyone wins.
In all other cases everyone loses.
That completes the specification of \elec.

Clearly \elec\ has a p-time winner problem, and by Theorem~\ref{t:upper-bounds}
we know that \escontrolmf{CCDV}
is in \pitwo. So what is left is to show that \escontrolmf{CCDV}
is \pitwo-hard.

Let $(\forall x_{\ell+1}, \ldots, x_{2\ell})(\exists x_{1},\ldots, x_{\ell})
[\psi(x_1,\ldots,x_{2\ell})]$ be an instance of $\qbfpi{2}$.
We construct an instance of \escontrolmf{CCDV} in the following way.
Let the candidate set $C$ consist of $p$ encoding $\psi$ and $\ell+1$ candidates
all lexicographically larger than $p$. So, the candidates in $C$ can be listed
in increasing lexicographic order as $p, c_1, \dots, c_{\ell + 1}$.
Let the collection of voters $V$ consist of one manipulator
and $2\ell$ nonmanipulators where
for each $i, 1 \le i \le \ell$, there is a voter
$v_i$ who votes $c_i > c_{\ell + 1} > p > \cdots$
and a voter $u_i$ who votes $c_i > p > c_{\ell + 1} > \cdots$.
Let the delete limit $k = 2\ell+1$
(any limit $\geq \ell$ will do)
and let the preferred candidate of the
chair be $p \in C$.

Suppose $(\forall x_{\ell+1}, \ldots, x_{2\ell})(\exists x_{1},\ldots, x_{\ell})
[\psi(x_1,\ldots,x_{2\ell})] \in \qbfpi{2}$.
Consider a vote $\widehat{v}$ for the manipulator.
If $\widehat{v}$ ranks $c_0$ first then the chair deletes
$v_i$ for all $i, 1 \leq i \leq \ell$ to make $p$ a winner.
If $\widehat{v}$ ranks $c_i$ first, for some $i, 1 \leq i \leq \ell$, 
and states $c_{\ell + 1} > c_0$, then the chair deletes
$\{u_1, \ldots, u_\ell\}$ to make $p$ a winner.
If $\widehat{v}$ ranks $c_i$ first, for some $i, 1 \leq i \leq \ell$, 
and states $c_0 > c_{\ell + 1}$, then the chair deletes
$\{v_1, \ldots, v_\ell\}$ to make $p$ a winner.
If $\widehat{v}$ ranks $c_{\ell + 1}$ first, 
then consider the assignment to $x_{\ell + 1}, \ldots, x_{2\ell}$ induced
by $\widehat{v}$ and 
fix an assignment to $x_1, \dots, x_{\ell}$ such that
$\psi(x_1,\dots,x_{2\ell})$ is true.
For each $i, 1 \le i \le \ell$, if $x_i$ is true in the assignment the chair
deletes $u_i$ and if $x_i$ is false the chair deletes
$v_i$. This will make $p$ a winner.

Conversely, fix an assignment to $x_{\ell + 1}, \ldots, x_{2\ell}$.
Set the manipulator vote $\widehat{v}$ so that it induces this assignment and 
so that $c_{\ell + 1}$ is ranked first. Consider the set of voters left after
the chair has deleted voters to make $p$ a winner. Note that this 
set must include $\widehat{v}$ and a set of voters that induces an 
assignment to $x_1, \ldots, x_\ell$ that makes $\psi$ true.~\end{proofs}

\begin{theorem}\label{DCDV-MF:complete}
There exists an election system, \elecprime, with a p-time winner problem,
such that
\scontrolmf{\elecprime}{DCDV} is \pitwo-complete.
\end{theorem}

\begin{proofs}
Let \elecprime\ be defined as $\elec$ in Theorem~\ref{CCDV-CF:complete} except
replace %
``everyone loses'' with ``everyone wins'' and
``everyone wins'' with ``everyone loses.''

Note that for every election $(C,V)$ and every candidate $p \in C$,
$p$ is an \elec\ winner of $(C,V)$ if and only if $p$ is not an
\elecprime\ winner of $(C,V)$. This immediately implies that
\scontrolmf{\elecprime}{DCDV} = \scontrolmf{\elec}{CCDV}.
The result follows from
Theorem~\ref{CCDV-MF:complete}.~\end{proofs}

\begin{theorem}\label{CCPV-CF:complete}
There exists an election system \elec, with a p-time winner problem,
such that \escontrolcf{CCPV-\vpairtetp-\vpair{\emptyset}{revoting}} are each
\sigmatwo-complete.\end{theorem}

\begin{proofs}
Let \elec\ be defined in the following way.
Given an election $(C,V)$, do the following.

If $\|C\|  = 1$ then the sole candidate wins.

If $\|C\| = 2$ then the lexicographically larger candidate
wins.

If $\|C\| \ge 3$, $\|V\| = 2\ell$, the candidates in $C$ listed in increasing lexicographic
order are $c_0, c_1, \dots, c_{\ell+1}$, and candidate $c_0$
encodes a boolean formula $\psi(x_1, \ldots, x_{2\ell})$,
then if for each $i, 1 \le i \le \ell$, 
there are exactly two voters with the same vote who rank $c_i$
first no one wins, else $c_{\ell+1}$ wins.

If $\|C\| \ge 3$, $\|V\| = 2\ell + 1$, the candidates in $C$ listed in increasing lexicographic
order are $c_0, c_1, \dots, c_{\ell+1}$, candidate $c_0$
encodes a boolean formula $\psi(x_1, \ldots, x_{2\ell})$,
and for each $i, 1 \le i \le \ell$, there are at
least two voters with the same vote who rank $c_i$ first, then do the following.
For each $i, 1 \le i \le \ell$, set $x_i$ to true if two voters with $c_i$
first both state $c_{\ell+1} > c_0$ and otherwise set $x_i$ to false.
Let $\widehat{v}$ be the unique vote that occurs three times or only once in $V$.
For each $i, 1 \le i \le \ell$, set $x_{\ell+i}$ to true if $\widehat{v}$
states $c_{i} > c_0$, else set $x_{\ell+i}$ to false.
If this is a satisfying assignment for $\psi$ then $c_0$ wins.

In all other cases, everyone loses.
That completes the specification of \elec.
    
Clearly \elec\ has a p-time winner problem, and
\escontrolcf{CCPV-\vpairtetp-\vpair{\emptyset}{revoting}}
are each in \sigmatwo\ by Theorem~\ref{t:upper-bounds}. So, what is left is to show
that \escontrolcf{CCPV-\vpairtetp-\vpair{\emptyset}{revoting}} are each
\sigmatwo-hard.

Let $(\exists x_1, \ldots, x_\ell)(\forall x_{\ell+1}, \ldots, x_{2\ell})
[\psi(x_1,\ldots,x_{2\ell})]$ be an instance of \qbfsigma{2}.
We construct an instance of \escontrolcf{CCPV-TE} in the following way.
Let the candidate set $C$ consist of $p$ encoding $\psi$ and
$\ell+1$ candidates lexicographically larger than $p$.
So, the candidates in $C$ can be listed in increasing lexicographic order
as $p, c_1, \dots, c_{\ell+1}$.
Let there be one manipulative voter, and let the nonmanipulators
consist of $2\ell$ pairs where
for each $i, 1 \le i \le \ell$, there are two voters $v_i$ and $v'_i$
with the same vote
$c_i > c_{\ell+1} > p > \cdots$
and two voters $u_i$ and $u'_i$ with the same 
vote $c_i > p > c_{\ell+1} > \cdots$.
Let the preferred candidate of the chair be $p \in C$.

If $(\exists x_1, \ldots, x_\ell)(\forall x_{\ell+1}, \ldots, x_{2\ell})
[\psi(x_1,\ldots,x_{2\ell})] \in \qbfsigma{2}$, fix an assignment to
$x_1, \dots, x_{\ell}$ such that
$(\forall x_{\ell+1}, \dots, x_{2\ell})[\psi(x_1, \dots, x_{2\ell})]$ is true.
The chair sets $V_1$ to consist of the manipulator and the subcollection of the voters
whose votes encode the assignment, i.e.,
for each $i, 1 \le i \le \ell$, if $x_i$ is true in the assignment the chair
adds $v_i$ and $v'_i$ to $V_1$ and if $x_i$ is false the chair adds
$u_i$ and $u'_i$ to $V_1$.
The chair puts the remaining voters from $V$ into $V_2$.
Note that the vote of the manipulator will be the unique vote
$\widehat{v}$ that occurs three times (if the manipulator votes for one
of the paired voters) or only once in $V_1$.
And no matter what assignment to
$x_{\ell + 1}, \ldots, x_{2\ell}$ is induced by $\widehat{v}$,
formula $\psi$ is satisfied and so $p$ is the unique winner of
$(C,V_1)$. Since $V_2$ consists $2\ell$ voters of the correct
form, no one wins $(C,V_2)$.
Only candidate $p$ participates in the runoff and so $p$ wins the
runoff. Note that this argument works for the ``TE'' and the ``TP'' models 
with or without revoting.

Conversely, if the chair can ensure that $p$ wins then there
exists a partition such that for all manipulations $p$ wins.
It is clear that the chair must partition the voters into
$(V_1, V_2)$ such that $\|V_1\| = 2\ell+1$ and $\|V_2\| = 2\ell$,
since otherwise there are no winners. Also, 
for each $i, 1 \leq i \leq \ell$, $V_2$ contains exactly two voters
with the same vote who rank $c_i$ first. It follows that $V_1$
contains the manipulator vote $\widehat{v}$ and that
for each $i, 1 \leq i \leq \ell$, $V_1$ contains exactly two nonmanipulators
with the same vote who rank $c_i$ first.
These $2\ell$ nonmanipulators
induce an assignment to $x_1, \ldots, x_\ell$. Fix this assignment.
Now fix an assignment to $x_{\ell+1}, \dots, x_{2\ell}$.
Set the manipulator vote $\widehat{v}$ so that it induces
this assignment. Since $p$ wins the runoff, this is a
satisfying assignment for $\psi$. It follows that for the
assignment to $x_1, \ldots, x_\ell$ that is induced by $V_1$,
it holds that
$(\forall x_{\ell+1}, \ldots, x_{2\ell}) [\psi(x_1,\ldots,x_{2\ell})]$ is true~\end{proofs}

\begin{theorem}\label{CCPV-MF:complete}
There exists an election system \elec, with a p-time winner problem,
such that \escontrolmf{CCPV-\vpairtetp} and \escontrolmfrevoting{CCPV-TE} are each \pitwo-complete.
\end{theorem}

\begin{proofs}
Let \elec\ be as defined in the proof of Theorem~\ref{CCPV-CF:complete}.
Then \elec\ has a p-time winner problem and by Theorem~\ref{t:upper-bounds} we know
that \escontrolmf{CCPV-\vpairtetp} and \escontrolmfrevoting{CCPV-TE} are each in \pitwo. So what is left is to show
that \escontrolmf{CCPV-\vpairtetp} and \escontrolmfrevoting{CCPV-TE} are each \pitwo-hard.
Below we describe the reduction for the ``TE''  case.
It is easy to see that the same reduction holds for the ``TP'' case. For
the ``TE'' case with revoting observe that the same reduction also holds since
in the runoff there will be at most two candidates and in 
election system \elec\ the votes do not affect who wins in that case.

Let $(\forall x_{\ell+1}, \ldots, x_{2\ell})(\exists x_{1}, \ldots, x_{\ell})
[\psi(x_1,\ldots,x_{2\ell})]$ be an instance of \qbfpi{2}. Our instance of
\escontrolmf{CCPV-TE} is exactly the instance of \escontrolcf{CCPV-TE} from the
proof of Theorem~\ref{CCPV-CF:complete}.
Note that the vote of the manipulator will always be the unique vote
$\widehat{v}$ that occurs three times or only once in $V$.
The same argument as in the proof of Theorem~\ref{CCPV-CF:complete} shows
that $(\forall x_{\ell+1}, \ldots, x_{2\ell})(\exists x_{1}, \ldots, x_{\ell})
[\psi(x_1,\ldots,x_{2\ell})] \in \qbfpi{2}$ if and only if the chair can
ensure that $p$ always becomes a winner by partitioning voters.~\end{proofs}

When revoting is allowed after the first round in the TP case,
and the manipulators go first, we find an interesting rise in complexity.

\begin{theorem}\label{CCPV-MF-rev:complete}
There exists an election system \elec, with a p-time winner problem,
such that \escontrolmfrevoting{CCPV-TP} is \pithree-complete.
\end{theorem}

\begin{proofs}
The election system, \elec, defined below will utilize the following
special candidates.

\begin{description} 
  \item[$\pair{1,\psi}{:}$] where $\psi$ is a boolean formula,
  which we refer to as a type-1 candidate.
  \item[$\pair{2,i,j}{:}$] where $i \in \naturalnumber$ and $j \in \{0,1\}$,
  which we refer to as a type-2 candidate.
  \item[$\pair{3,i,j}{:}$] where $i \in \naturalnumber$ and $j \in \{0,1\}$,
  which we refer to as a type-3 candidate.
  \item[$\pair{4,i}{:}$] where $i \in \naturalnumber$, which we refer to as
  a type-4 candidate. %
\end{description}

Let \elec\ be defined in the following way.

\noindent
Given an election $(C,V)$:

If $C$ consists of %
one type-1 candidate encoding $\psi(x_1,\dots,x_{3\ell})$,
$2\ell$ type-2 candidates
$\pair{2, 1, 0}, \pair{2,1,1}, \dots, \pair{2, \ell, 0}, \pair{2,\ell,1}$,
$2\ell$ type-3 candidates $\pair{3, 1, 0}, \pair{3,1,1}, \dots,
\pair{3, \ell, 0}, \pair{3,\ell,1}$, and $\ell+2$ type-4 candidates
$\pair{4,1}, \dots, \pair{4,\ell+2}$, then do the following.
\begin{itemize}

\item If $\|V\| = 2\ell + 1$ and for each $i, 1 \le i \le \ell$, there are
at least two voters with the same vote who rank $\pair{4,i}$ first,
then we have $3\ell+2$ winners consisting of
$\pair{1,\psi}$, $\pair{4,1}, \dots, \pair{4,\ell+1}$, and
$2\ell$ candidates determined in the following way.
Let $\widehat{v}$ be the unique vote that occurs three times or only once
in $V$.
For each $i, 1 \le i \le \ell$, $\pair{2,i,1}$ is a winner
if $\widehat{v}$ states $\pair{4,i} > \pair{1,\psi}$
and otherwise $\pair{2,i,0}$ is a winner.
For each $i, 1 \le i \le \ell$, $\pair{3,i,1}$ is a winner
if two voters who rank $\pair{4,i}$ first both state
$\pair{4,\ell+1} > \pair{1,\psi}$ and otherwise $\pair{3,i,0}$ is a winner.

\item If $\|V\| = 2\ell$ then if for each $i, 1 \le i \le \ell$, there are
at least two voters with the same vote who rank $\pair{4,i}$ first, no one wins,
else $\pair{4,\ell+2}$ wins.
\end{itemize}

If $C$ consists of one type-1 candidate
encoding $\psi(x_1,\dots,x_{3\ell})$,
$\ell$ type-2 candidates of the form
$\pair{2,1,\star}, \dots, \pair{2,\ell,\star}$
(where $\star \in \{0,1\}$),
$\ell$ type-3 candidates of the form
$\pair{3,1,\star}, \dots, \pair{3,\ell,\star}$
(where $\star \in \{0,1\}$),
and $\ell+1$ type-4 candidates
$\pair{4,1}, \dots, \pair{4,\ell+1}$,
$\|V\| = 4\ell + 1$,
and there is a unique vote $\widehat{v}'$ that occurs
three times or only once in $V$,
then do the following.
For each $i, 1 \le i \le \ell$, set $x_{i}$ to true if
$\pair{2,i,1}$ is in $C$ and to false if $\pair{2,i,0}$ is in $C$,
set $x_{\ell+i}$ to true if 
$\pair{3,i,1}$ is in $C$ and to false if $\pair{3,i,0}$ is in $C$, and
set $x_{2\ell+i}$ to true if $\widehat{v}'$ states $\pair{4,i} > \pair{1,\psi}$
and else set $x_{2\ell+i}$ to false.
If this is a satisfying assignment for formula $\psi$, then $\pair{1,\psi}$
wins. Otherwise, everyone loses.

Else, everyone loses.

\noindent
That completes the specification of \elec.

Clearly \elec\ has a p-time winner problem, and 
\escontrolmfrevoting{CCPV-TP} is in \pithree\ by Theorem~\ref{t:upper-bounds}.
So, what is left to show is that \escontrolmfrevoting{CCPV-TP}
is \pithree-hard.

Let $(\forall x_{1}, \ldots, x_{\ell})(\exists x_{\ell+1}, \dots,
x_{2\ell})(\forall x_{2\ell+1}, \dots, x_{3\ell})[\psi(x_1,\dots,x_{3\ell})]$
be an instance of $\qbfpi{3}$.
We construct an instance of \escontrolmfrevoting{CCPV-TP} in the following way.
Let the candidate set $C$ consist of
one type-1 candidate encoding $\psi$,
$2\ell$ type-2 candidates
$\pair{2,1,0}, \pair{2,1,1}, \dots, \pair{2,\ell,0}, \pair{2,\ell,1}$,
$2\ell$ type-3 candidates
$\pair{3,1,0}, \pair{3,1,1}, \dots, \pair{3,\ell,0}, \pair{3,\ell,1}$,
and $\ell+2$ type-4 candidates
$\pair{4,1}, \dots, \pair{4,\ell+2}$.
Let there be one manipulator and $4\ell$ nonmanipulators
where for each $i, 1 \le i \le \ell$, there are two
voters $v_i$ and $v'_i$ with the same vote
$\pair{4,i} > \pair{4,\ell+1} > \pair{1,\psi} > \cdots$
and two voters $u_i$ and $u'_i$ with the same vote
$\pair{4,i} > \pair{1,\psi} > \pair{4,\ell+1} > \cdots$.
Let the preferred candidate of the chair be $\pair{1,\psi} \in C$.

Suppose $(\forall x_{1}, \dots, x_{\ell})(\exists x_{\ell+1}, \dots,
x_{2\ell})(\forall x_{2\ell+1}, \dots, x_{3\ell})[\psi(x_1,\dots,x_{3\ell})]
\in \qbfpi{3}$.
Consider a first-round vote $\widehat{v}$ for the manipulator, and view
it as an assignment to $x_1, \dots, x_{\ell}$ where for each $i, 1 \le i \le \ell$,
if $\widehat{v}$ states $\pair{4,i} > \pair{1,\psi}$ then $x_i$ is true and otherwise
$x_i$ is false. Using this assignment, set an assignment to
$x_{\ell+1}, \dots, x_{2\ell}$ such that
$(\forall x_{2\ell+1}, \dots, x_{3\ell})\psi(x_1,\dots,x_{3\ell})$ is true. 
The chair sets $V_1$ to consist of the manipulator and for each $i, 1 \le i \le \ell$,
if $x_{\ell+i}$ is true in the assignment the chair adds $v_i$ and $v'_i$ to $V_1$ and
if $x_{\ell+i}$ is false the chair adds $u_i$ and $u'_i$ to $V_1$. The chair puts the
remaining voters from $V$ into $V_2$. Note that $\widehat{v}$ will be the unique vote
that occurs three times or only once in $V_1$. Notice that the type-2 and type-3
candidates that proceed to the runoff ``hold'' the abovementioned assignments to $x_1, \dots, x_{\ell}$
and $x_{\ell+1}, \dots, x_{2\ell}$ respectively
(since $\pair{2,i,1}$ proceeds to the runoff if and only if $x_i$ is true,
$\pair{2,i,0}$ proceeds to the runoff if and only if $x_i$ is false,
$\pair{3,i,1}$ proceeds to the runoff if and only if $x_{\ell+i}$ is true, and
$\pair{3,i,0}$ proceeds to the runoff if and only if $x_{\ell+i}$ is false).
And that no matter what assignment to
$x_{2\ell+1}, \dots, x_{3\ell}$ is induced by the second-round vote $\widehat{v}'$ of the manipulator,
formula $\psi$ is true and so $\pair{1,\psi}$ wins.

Conversely, suppose that for all first-round manipulator votes there exists a
partition such that for all second-round manipulator votes
$\pair{1,\psi}$ wins.
Fix a first-round manipulator vote $\widehat{v}$, and let $(V_1,V_2)$
be a partition such that $\pair{1,\psi}$ wins regardless of the second-round
vote of the manipulator.
It is clear that $\|V_1\| = 2\ell + 1$ and $\|V_2\| = 2\ell$ (without loss of generality),
and that for each $i, 1 \le i \le \ell$, $V_2$ contains exactly two
voters with the same vote who rank $\pair{4,i}$ first. It follows that
the first-round manipulator vote $\widehat{v}$ is the unique vote that
occurs three times or only once in $V_1$ and that
for each $i, 1 \le i \le \ell$,
$V_1$ contains two voters with the same vote who rank
$\pair{4,i}$ first.

Fix an assignment to $x_1, \dots, x_{\ell}$ and consider the first-round manipulator
vote $\widehat{v}$ where for each $i, 1 \le i \le \ell$, if $x_i$ is true then
$\widehat{v}$ states $\pair{4,i} > \pair{1,\psi}$ and so $\pair{2,i,1}$ proceeds to the runoff,
and if $x_i$ is false then $\widehat{v}$ states $\pair{1,\psi} > \pair{4,i}$ and so $\pair{2,i,0}$ proceeds to the runoff.
Since we know that there exists a partition $(V_1,V_2)$ where $\pair{1,\psi}$ wins the runoff, we know that
for each $i, 1 \le i \le \ell$, $\pair{3, i, 1}$ proceeds to the runoff
if $v_i$ and $v'_i$ are in $V_1$ and otherwise $\pair{3, i, 0}$ does. We can
view this as an assignment to $x_{\ell+1}, \dots, x_{2\ell}$ where for each
$i, 1 \le i \le \ell$, if $\pair{3,i,1}$ proceeds to the runoff then $x_i$ is true
and if $\pair{3,i,0}$ proceeds to the runoff then $x_i$ is false.
Now fix an assignment to $x_{2\ell+1}, \dots, x_{3\ell}$ and set
the second-round manipulator vote $\widehat{v}'$ so that it induces this
assignment. Since $\pair{1,\psi}$ wins the runoff, $\psi$ is true for
the assignment to $x_1, \dots, x_{\ell}, x_{\ell+1}, \dots, x_{2\ell}, x_{2\ell+1}, \dots, x_{3\ell}$.
It follows that $(\forall x_1, \dots, x_{\ell})(\exists x_{\ell+1}, \dots, x_{2\ell})
(\forall x_{2\ell+1}, \dots, x_{3\ell})[\psi(x_1, \dots, x_{3\ell})] \in \qbfpi{3}$.~\end{proofs}

\section{Specific Systems}\label{app:specific}

In some of the proofs in this section,
we use the notation $\scoresub{(C,V)}{a}$
to denote the score of candidate $a$ in election $(C,V)$.
When it is clear from context, we may leave out $C$, $V$, or both.

\subsection{Plurality}

\begin{theorem}
For plurality elections, the following hold.

\begin{enumerate}
 \item {\rm M+{\vpaircd{}C\vpairad{}V}} are each in $\p$.
 \item {\rm \vpaircd{}C\vpairad{}V-\vpaircfmf} are each in $\p$.
\end{enumerate}

\end{theorem}

\begin{proofs}
For the constructive cooperative and the destructive competitive cases it is clear that
the manipulators should all vote for $p$.

For the destructive cooperative and the constructive competitive cases the optimal
action for the manipulators is to all vote for the same highest-scoring
candidate in $C - \{p\}$.

In all cases we can determine if the chair can be successful by assuming the manipulators
vote as above and using the corresponding
p-time algorithm for control from Bartholdi, Tovey, and Trick~\shortcite{bar-tov-tri:j:control}
(for the constructive cases) or from
Hemaspaandra, Hemaspaandra, and Rothe~\shortcite{hem-hem-rot:j:destructive-control}
(for the destructive cases), modified in the obvious way for the nonunique-winner
case (see Observation~\ref{obs:winner-model}).~\end{proofs}

For the remaining proofs in this section,
given an election $(C,V)$ containing $k$ manipulators,
we say that a candidate $r$ is a \emph{rival} of
$p$ if $r$ can beat $p$ pairwise, i.e.,
if $\scoresub{\{p,r\}}{r} + k > \scoresub{\{p,r\}}{p}$.

\begin{lemma}\label{lem:plurality-ccpv}
If there exists a partition such that $p$ is an overall winner in the
``TE'' model when all manipulators vote for the same highest-scoring
rival $r$ and put $p$ last, then there exists a partition
such that $p$ is always an overall winner.
\end{lemma}

\begin{proofs}
Given an election $(C,V)$ where $V$ contains $k$ manipulators, a candidate $p \in C$, and
a candidate $r \in C-\{p\}$ such that $\scoresub{\{p,r\}}{r} + k > \scoresub{\{p,r\}}{p}$,
we do the following.

Let $(V_1,V_2)$ be a partition such that $p$ is an
overall winner when all manipulators vote for $r$ and put $p$ last.
Let $k_1$ be the number of manipulators in $V_1$, 
let $k_2$ be the number of manipulators in $V_2$, 
let $\ell_1$ be the number of nonmanipulator votes for $r$ in $V_1$, and
let $\ell_2$ be the number of nonmanipulator votes for $r$ in $V_2$.
Without loss of generality assume that $p$ is 
the unique winner of
$(C,V_1)$ when all manipulators vote for $r$.

Now we will construct a new partition $(\widehat{V}_1,\widehat{V}_2)$
that will work regardless of how the manipulators vote.
Let $\widehat{V}_2$ consist of $\ell_2$ nonmanipulator votes for $r$,
$\scoresub{V_2}{p}$ nonmanipulator votes for $p$,
for every
rival $\widehat{r} \neq r$, $\mymin{\ell_2,\score{\widehat{r}}}$ votes
for $\widehat{r}$, for every 
nonrival %
$c \neq p$ all the
nonmanipulator votes for $c$, and $k_2$ manipulators.
Let $\widehat{V}_1 = V-\widehat{V}_2$.

We first show that $p$ is always the unique winner of $(C,\widehat{V}_1)$.
We know that
$\scoresub{\widehat{V}_1}{r} + k_1 = \ell_1 + k_1 < \scoresub{V_1}{p} =
\scoresub{\widehat{V}_1}{p}$.
For every nonrival $c \neq p$,
$\scoresub{\widehat{V}_1}{c} + k_1 = k_1 < \scoresub{\widehat{V}_1}{p}$.
Finally, for every rival $\widehat{r} \neq r$,
$\score{\widehat{r}} \leq \score{r} = \ell_1 + \ell_2$, and so
$\scoresub{\widehat{V}_1}{\widehat{r}}  \leq \ell_1$, which implies
that $\scoresub{\widehat{V}_1}{\widehat{r}} + k_1 \leq \ell_1 + k_1
< \scoresub{\widehat{V}_1}{p}$.
It follows that $p$ is always the unique winner of $(C,\widehat{V}_1)$.

So the only way in which $p$ can be precluded from winning the runoff
is if there exist a manipulation and a rival $\widehat{r}$ of $p$
such that $\widehat{r}$ is the unique winner of $(C,\widehat{V}_2)$.
Then $\ell_2 + k_2 > \score{c}$ for every nonrival $c \neq p$, and
$\ell_2 + k_2 > \scoresub{\widehat{V}_2}{p} =\scoresub{V_2}{p}$.
Now consider $(C,V_2)$ and let all manipulators vote for $r$.
Then the score of $r$ in $(C,V_2)$ (after the manipulation) is $\ell_2 + k_2$,
and $r$ is the unique winner of $(C,V_2)$.
Then $p$ is not an overall
winner of $(C,V)$ when all manipulators vote for $r$, which
contradicts our assumption. 

It follows that $p$ is always a winner of
$(\widehat{V}_1, \widehat{V}_2)$.~\end{proofs}

\begin{theorem}\scontrolcf{plurality}{CCPV-TE} is in \p.\end{theorem}

\begin{proofs}
Given an election $(C,V)$ and a preferred candidate of the chair $p \in C$,
$p$ can be made a winner if and only if there exists a partition $(V_1,V_2)$
such $p$ is always an overall winner. %

If no rivals of $p$ exist, then clearly control is possible if and only if
$C = \{p\}$ or there is at least one vote for $p$
(in the latter case, let $V_1$ consist of one voter for $p$).

Otherwise, let $r$ be a highest-scoring rival of $p$.
It is immediate from Lemma~\ref{lem:plurality-ccpv} that control is possible
if and only if there exists a partition such that 
$p$ wins when all manipulators vote for $r$ and put $p$ last.
This can be determined by running the polynomial-time
algorithm for plurality-CCPV-TE from~\cite{hem-hem-rot:j:destructive-control},
modified in the obvious way for the nonunique-winner case
(see Observation~\ref{obs:winner-model}).~\end{proofs}

\begin{theorem}$\scontrolcf{{\rm plurality}}{CCPV-TE} = \scontrolmf{{\rm plurality}}{CCPV-TE}.$
\end{theorem}

\begin{proofs}
It immediately follows from the definition that
$\scontrolcf{{\rm plurality}}{CCPV-TE} \subseteq \scontrolmf{{\rm plurality}}{CCPV-TE}.$

Now suppose that ``MF'' control is possible. Then
for all manipulations there exists a partition such that the preferred
candidate $p$ wins.
Then either no rival to $p$ exists, in which case ``CF'' control is possible
since either $p$ is the only candidate or there exists at least one vote for
$p$.
When a rival $r$ to $p$ exists, control is certainly possible when all 
the manipulators vote for $r$ and put $p$ last.
By Lemma~\ref{lem:plurality-ccpv} we know that then
there exists a partition where $p$ is always a winner, so ``CF'' control is
possible.~\end{proofs}
 
\begin{corollary}\scontrolmf{plurality}{CCPV-TE} is in \p.\end{corollary}

\begin{theorem}\mcontrol{plurality}{DCPV-TE} is in \p.\end{theorem}

\begin{proofs}
Given an election $(C,V)$ and a despised candidate of the chair $p \in C$, we
can determine in polynomial time if $p$ can be precluded from winning
by partitioning voters as follows.
If there are no manipulators, run the polynomial-time
algorithm for plurality-DCPV-TE from~\cite{hem-hem-rot:j:destructive-control},
modified in the obvious way for the nonunique-winner
case (see Observation~\ref{obs:winner-model}).

So, let $k > 0$ denote the number of manipulators in $V$.
If there exists a rival $r$ to $p$ (i.e., a candidate that can beat
$p$ pairwise, i.e., a candidate for which
$\scoresub{\{p,r\}}{p} < \scoresub{\{p,r\}}{r} + k$), then
control is possible:  Let $V_2$ consist of one manipulator and let
all manipulators vote for $r$.

If there are no rivals, we must ensure that $p$ doesn't make it to
the runoff.
It is easy to see that this can be done if and only if we are in
one of the following two cases.
\begin{enumerate}
\item There are at least two candidates, $c$ is a highest-scoring
candidate in $C - \{p\}$, and  $\score{p} \leq \score{c} + k$.
(Have all manipulators vote for $c$ and use partition $(V,\emptyset)$.)
\item There are at least three candidates, $c$ and $d$ are two
highest-scoring candidates in $C - \{p\}$, and
$\score{p} \le \score{c}+\score{d}+k$.
(Have $V_1$ consist of $\min(\score{p},\score{c})$ votes for $p$
and all votes for $c$. The remaining votes, including all manipulators,
who will vote for $d$, will be in $V_2$.)
\end{enumerate}~\end{proofs}

\begin{lemma}\label{lem:dcpv-te:plur}
If there exists a partition of voters such that $p$ is not a plurality
winner in the ``TE'' model when all
manipulators vote for $p$, then there exists a partition such that
$p$ can never be made a plurality winner by the manipulators.\end{lemma}

\begin{proofs}
Given an election $(C,V)$ and a candidate $p \in C$, we do the following.

Let $(V_1,V_2)$ be a partition such that $p$ is not a winner when all
manipulators vote for $p$.
If $p$ can never be made a winner by the manipulators in this partition
then we are done.
So, suppose there exists a manipulation such that $p$ is an overall winner
(with the partition $(V_1,V_2)$).
Without
loss of generality assume that $p$ is 
the unique winner of $(C,V_1)$.
Then $p$ is also the unique winner in $(C,V_1)$ if all manipulators vote for
$p$. However, since $p$ is not an overall winner if all manipulators vote for $p$ 
there is a candidate $c \neq p$ such that if all manipulators 
vote for $p$, $c$ is the unique winner of $(C,V_2)$ and $c$ is the unique
winner of the runoff $(\{p,c\},V)$.

Now move all manipulators from $V_2$ to $V_1$. Note that $c$ remains the
unique winner of $(C,V_2)$ and that $c$ is always the unique winner of
$(\{p,c\},V)$. It follows that in this new partition, $p$ is never a winner,
no matter what the manipulators do.~\end{proofs}

Lemma~\ref{lem:dcpv-te:plur} implies that \scontrolcf{plurality}{DCPV-TE} is in \p,
since control is possible if and only if control is possible when
all manipulators vote for $p$. This can
be checked using the polynomial-time
algorithm for \scontrol{plurality}{DCPV-TE} from Hemaspaandra, Hemaspaandra, and Rothe~\shortcite{hem-hem-rot:j:destructive-control},
modified in the obvious way for the nonunique-winner case
(see Observation~\ref{obs:winner-model}).

\begin{theorem}\scontrolcf{plurality}{DCPV-TE} is in \p.\end{theorem}

We will now show that
Lemma~\ref{lem:dcpv-te:plur} also implies that
\scontrolmf{plurality}{DCPV-TE} is in \p.

\begin{theorem}\label{thm:plur:dcpv-mf-te}
\scontrolmf{plurality}{DCPV-TE} is in \p.\end{theorem}

\begin{proofs}
Given an election $(C,V)$ and a despised candidate of the chair $p \in C$,
we will show that we can
determine in polynomial time if $p$ can be precluded from winning by
partitioning voters.

As in the ``{\rm CF}'' case we will use Lemma~\ref{lem:dcpv-te:plur} to
show that control is possible if and only 
if there exists
a partition such that $p$ is precluded from winning when all manipulators vote
for $p$. This also implies that
\scontrolcf{plurality}{DCPV-TE} = \scontrolmf{plurality}{DCPV-TE}.

It immediately follows from the
definition that if the instance of
\scontrolmf{plurality}{DCPV-TE}  is positive,
then there exists a partition such that $p$ is not a winner when all
manipulators vote for $p$.

For the other direction, by Lemma~\ref{lem:dcpv-te:plur} if there exists a
partition such that $p$ is not a winner when all the manipulators vote for
$p$, then there exists a partition $(V_1,V_2)$ such that $p$ can never be made
a winner by the manipulators. This implies that no matter what the manipulators
do, there exists a partition (in fact, always the same partition)
such that $p$ is not a winner. This then implies that the instance of
\scontrolmf{plurality}{DCPV-TE} is positive.~\end{proofs}

\subsection{Condorcet}

\begin{theorem}
For Condorcet elections, the following hold.

\begin{enumerate}
 \item {\rm M+{\vpaircd{}C\vpairad{}C}} are each in $\p$.
 \item {\rm M+{DC\vpairad{}V}} are both in $\p$.
 \item {\rm {\vpaircd{}C\vpairad{}C}-\vpaircfmf} are each in $\p$.
 \item {\rm {DC\vpairad{}V-\vpaircfmf}} are both in $\p$.
\end{enumerate}

\end{theorem}

\begin{proofs}
For the constructive cooperative and the destructive competitive cases it is
clear that the manipulators should all vote for $p$.

For the destructive cooperative and the constructive competitive cases the
optimal action for the manipulators is to rank $p$ last.

In all cases we can determine if the chair can be successful by assuming the
manipulators vote as above and using the corresponding p-time algorithm for
control from Bartholdi, Tovey, and Trick~\shortcite{bar-tov-tri:j:control}
(for the constructive cases) or from Hemaspaandra, Hemaspaandra,
and Rothe~\shortcite{hem-hem-rot:j:destructive-control}
(for the destructive cases), modified in the obvious way for the nonunique-winner
case (see Observation~\ref{obs:winner-model}).~\end{proofs}

We now prove the Condorcet partition cases. Since Condorcet winners are
always unique, the ``TE'' and ``TP'' cases coincide and so we will leave
out this notation, following~\cite{hem-hem-rot:j:destructive-control}.

\begin{theorem}\label{thm:con-m+ccpc}
\mcontrol{Condorcet}{\vpaircd{}C\vpair{PC}{RPC}} are each in \p.
\end{theorem}

\begin{proofs}
Given an election $(C,V)$ and a preferred candidate of the chair $p \in C$,
we can determine in polynomial time if $p$ can be made a winner by
partitioning of candidates and by runoff partitioning of candidates as follows.

For the constructive cases we do the following.
Since Condorcet elections satisfy both WARP and unique-WARP,
we know from Theorems~\ref{thm:warp-one-part} and~\ref{thm:warp-two-part} that control is possible
if and only if control is possible using partition $(C-\{p\},\{p\})$.
Set all manipulators to rank $p$ first.
Rank the candidates that do not beat $p$ pairwise next in all 
manipulator votes (in any order).
Then, as long as there exists an unranked candidate $c$
that can never be a Condorcet winner in $(C-\{p\}, V)$,
rank $c$ next in all manipulator votes.

Let $\widehat{C}$ be the set of candidates not yet ranked by the
manipulators. Notice that every $c \in \widehat{C}$
beats $p$ pairwise, and every $c \in \widehat{C}$ can become
a Condorcet winner in $(\widehat{C},V)$ (and thus also in 
$(C,V)$).

So, to determine if control is possible, we must determine if the
manipulators can vote in such a way that there is no Condorcet
winner in $(\widehat{C}, V)$, i.e.,
$\forall c \in \widehat{C}\, \exists c' \in \widehat{C}$ such that
$c'$ ties-or-beats $c$ pairwise.

For $\|V\|$ even, assume that there are at least two candidates in
$\widehat{C}$
and for $\|V\|$ odd, assume there are at least three candidates in
$\widehat{C}$
(otherwise there will always be Condorcet winners).
We have the following cases, depending on whether or not
there is a Condorcet winner in $(\widehat{C},V)$ before the manipulators vote
and depending on  the parity of $\|V\|$.
Let $k \geq 1$ denote the number of manipulators in $V$.

\begin{enumerate}
\item If there exists a Condorcet winner and $\|V\|$ is even, then let $c$ be the
Condorcet winner, and let $d \in \widehat{C}-\{c\}$. It is easy to see that each of the
manipulators can vote $c > d > \widehat{C}-\{c,d\}$ or $d > c > \widehat{C}-\{c,d\}$ in
such a way that $c$ ties $d$ pairwise. So, $c$ is no longer a Condorcet winner and
no other candidate becomes a Condorcet winner, since $c$ ties-or-beats every
other candidate pairwise.

\item If there exists a Condorcet winner and $\|V\|$ is odd, then let $c$ be the
Condorcet winner, and let $a,b \in \widehat{C}-\{c\}$ be such that $a$ ties-or-beats $b$ pairwise.
Have $\lceil k/2 \rceil$ manipulators vote $a > b > c > \widehat{C}-\{a,b,c\}$ and
$\lfloor k/2 \rfloor$ manipulators vote $b > c > a > \widehat{C}-\{a,b,c\}$.
After this manipulation, $b$ beats
$c$ pairwise, $a$ beats $b$ pairwise, and $c$ beats every candidate
in $\widehat{C} - \{b,c\}$ pairwise.

\item If there is no Condorcet winner and $\|V\|$ is even, then have $\lfloor k/2 \rfloor$ manipulators vote $\widehat{C}$ (i.e., the candidates
in $\widehat{C}$ in some fixed order)
and $\lfloor k/2 \rfloor$ manipulators vote $\overleftarrow{\widehat{C}}$
(i.e., the candidates in $\widehat{C}$ in reverse order).
 When $k$ is odd, let the remaining
manipulator vote arbitrarily.
It is clear that no Condorcet winners are created by the manipulators.

\item If there is no Condorcet winner and $\|V\|$ is odd, then we have the following cases.

\begin{enumerate}

\item If $k$ is even, then have $k/2$ manipulators vote $\widehat{C}$ and 
the remaining $k/2$ manipulators vote $\overleftarrow{\widehat{C}}$.

\item If $k$ is odd and there is no weak Condorcet winner
(a weak Condorcet winner is a candidate that ties-or-beats 
every other candidate pairwise),
then have $\lfloor k/2 \rfloor$ manipulators vote $\widehat{C}$
and $\lfloor k/2 \rfloor$ manipulators vote $\overleftarrow{\widehat{C}}$. Let the remaining
manipulator vote arbitrarily. It is clear that no Condorcet winner is created by the manipulators.

\item If $k$ is odd and there exists a weak Condorcet winner,
then let $c$ be a weak Condorcet winner and let $a$ be a candidate
such that $a$ ties $c$ pairwise.  We have the following two cases.
\begin{enumerate}

\item If for all $b \in \widehat{C}-\{a,c\}$, $a$ beats $b$ pairwise and $c$ beats $b$ pairwise, then have
$\lceil k/2 \rceil$ manipulators vote $\widehat{C}-\{a,c\} > a > c$ and have the remaining $\lfloor k/2 \rfloor$ manipulators
vote $c > \widehat{C}-\{a,c\} > a$. So, now $a$ beats $c$ pairwise, and for all $b \in \widehat{C}-\{a,c\}$,
$c$ beats $b$ pairwise and $b$ beats $a$ pairwise,
and thus there is still no Condorcet winner.

\item Otherwise, there exists a candidate $b \in C-\{a,c\}$ such that it
is not the case that $a$ and $c$ both beat $b$ pairwise.
Suppose there are at least three manipulators, and set their votes in the
following way. (If there is only one manipulator, then since each candidate in $\widehat{C}$ can become
a Condorcet winner, all candidates in $\widehat{C}$ tie pairwise.
And so there is always a Condorcet winner after manipulation.)
\begin{enumerate}
\item
If $a$ does not beat $b$ pairwise, then
let $\lfloor k/3 \rfloor$ manipulators vote
$c > b > a > \widehat{C}-\{a,b,c\}$,
 $\lfloor k/3 \rfloor$ manipulators vote $b > a > c > \widehat{C}-\{a,b,c\}$, and
$\lfloor k/3 \rfloor$ manipulators vote
$a > c > b > \widehat{C}-\{a,b,c\}$. 
Note that $a$ beats $c$ pairwise, $b$ beats $a$ pairwise, and
$c$ beats every candidate in $\widehat{C} - \{a,c\}$ pairwise, so
there is no Condorcet winner.
If two manipulators remain, then have one vote $\widehat{C}$ and
the other vote $\overleftarrow{\widehat{C}}$. Otherwise, if a single manipulator remains, since $a$ beats $c$ pairwise
after the manipulators act as above, when the one remaining manipulator votes $c > \cdots$, no Condorcet winner is
created.
\item
If $a$ beats $b$ pairwise, then $c$ does not beat $b$ pairwise.
It follows that $c$ ties $b$ pairwise. Now switch candidates $a$ and $b$, 
and we are in the previous case.
\end{enumerate}
\end{enumerate}
\end{enumerate}
\end{enumerate}

For the destructive cases, since Condorcet elections satisfy unique-WARP, the chair
cannot, by partitioning of candidates or by runoff partitioning of candidates, cause a
candidate that is a unique winner to no longer be a unique winner~\cite{hem-hem-rot:j:destructive-control}.
This implies that
control is possible if and only if the manipulators can vote so that
$p$ is not a winner in $(C,V)$.
It is immediate that the optimal action for the manipulators is to put $p$
last.~\end{proofs}

\begin{theorem}\label{thm:con-ccpc-cfmf}
\scontrol{Condorcet}{\vpaircd{}C\vpair{PC}{RPC}-\vpaircfmf} are each in \p.
\end{theorem}

\begin{proofs}
Given an election $(C,V)$ and a preferred candidate of the
chair $p \in C$, we can determine in polynomial time if $p$ can be made
a winner by partitioning candidates and by runoff partitioning
of candidates as follows.

For the constructive cases,
since Condorcet elections satisfy both WARP and unique-WARP,
we know from Theorems~\ref{thm:warp-one-part} and~\ref{thm:warp-two-part} 
(which each apply only to the TE model, but since the Condorcet 
election system never has more than one winner, for Condorcet elections
TE and TP are in effect identical)
that  control
is possible if and only if control is possible using partition $(C-\{p\},\{p\})$.
The manipulators can preclude $p$ from winning if and only if there is a candidate
$c \neq p$ that can be made to uniquely win $(C-\{p\},V)$ and ties-or-beats $p$ pairwise.
This
can easily be checked by having all manipulators vote for $c$.

For the destructive cases, since Condorcet elections satisfy unique-WARP,
the chair cannot, by partitioning of candidates or by runoff partitioning
of candidates, cause a candidate that is a unique winner to no longer be
a unique winner~\cite{hem-hem-rot:j:destructive-control}.
This implies that control
is possible if and only if the manipulators cannot vote so that $p$ becomes a
winner in $(C,V)$. It is immediate that the optimal action for the manipulators
is to vote for $p$.~\end{proofs}

\begin{theorem}\label{thm:con-m+dcpv}
\mcontrol{Condorcet}{DCPV} is in \p.\end{theorem}

\begin{proofs}
Given an election $(C,V)$ and a despised candidate of the chair
$p \in C$, we can determine in polynomial time if $p$ can be
precluded from winning by partitioning voters as follows.

If there exists a candidate $r \in C-\{p\}$ such that when
all manipulators rank $p$ last, $r$ ties-or-beats $p$ pairwise,
then control is possible by having all manipulators rank $p$ last and
using partition $(V,\emptyset)$.

If no such candidate exists, the only way to ensure that $p$ is not a winner
is to ensure that $p$ does not participate in the runoff.
Suppose there exists a partition and a manipulation such that $p$ is not
a unique winner of either subelection. If in this partition we set all manipulators
to rank $p$ last, $p$ still does not win either subelection.
So, we can check whether we are in this case by having all manipulators rank
$p$ last, and then use the polynomial-time algorithm for
\scontrol{Condorcet}{DCPV} from~\cite{hem-hem-rot:j:destructive-control}, modified
in the obvious way for the nonunique-winner case
(see Observation~\ref{obs:winner-model}).~\end{proofs}

Below we state a lemma analogous to Lemma~\ref{lem:dcpv-te:plur}, but for Condorcet elections.

\begin{lemma}\label{lem:con:dcpv}
If there exists a partition of voters such that $p$ is not a Condorcet
winner when all manipulators vote for $p$, then there exists a partition
such that $p$ can never be made a winner by the manipulators.\end{lemma}

\begin{proofs}
Given an election $(C,V)$ and a candidate $p \in C$, we do the following.

Let $(V_1,V_2)$ be a partition such that $p$ is not a winner when all
manipulators vote for $p$. So, either there exists a candidate $r \in C-\{p\}$
such that $r$ ties-or-beats $p$ pairwise when all manipulators vote for $p$, or $p$
is not a unique winner of either subelection.

In the former case the partition $(V,\emptyset)$ will always work, and in the
latter case it is clear to see that there is no way for the manipulators to
make $p$ a unique winner of either subelection, so we are done.~\end{proofs}

Lemma~\ref{lem:con:dcpv} implies that
\scontrolcf{Condorcet}{DCPV} is in \p,
since control is possible if and only if control
is possible when all manipulators vote for $p$. This can be checked using
the polynomial-time algorithm from~\cite{hem-hem-rot:j:destructive-control},
modified in the obvious way for the nonunique-winner case
(see Observation~\ref{obs:winner-model}).

\begin{theorem}\label{thm:con-dcpv-cf}
\scontrolcf{Condorcet}{DCPV} is in \p.\end{theorem}

A similar argument as in the proof of Theorem~\ref{thm:plur:dcpv-mf-te}
shows that Lemma~\ref{lem:con:dcpv} above also implies that the corresponding
``MF'' case is also in \p.

\begin{theorem}\label{thm:con-dcpv-mf}
\scontrolmf{Condorcet}{DCPV} is in \p.\end{theorem}

\subsection{Approval}

\begin{theorem}
For approval elections, the following hold.

\begin{enumerate}
 \item {\rm M+{\vpaircd{}C\vpairad{}C}} are each in $\p$.
 \item {\rm M+{DC\vpairad{}V}} are both in $\p$.
 \item {\rm {\vpaircd{}C\vpairad{}C-\vpaircfmf}} are each in $\p$.
 \item {\rm {DC\vpairad{}V-\vpaircfmf}} are each in $\p$.
\end{enumerate}

\end{theorem}

\begin{proofs}
For the constructive cooperative and the destructive competitive cases it is
clear that the manipulators should all approve of only $p$.

For the destructive cooperative and the constructive competitive cases the
optimal action for the manipulators is approve of all candidates except
$p$.

In all cases we can determine if the chair can be successful by assuming the
manipulators vote as above and using the corresponding p-time algorithm for
control from Bartholdi, Tovey, and Trick~\shortcite{bar-tov-tri:j:control}
(for the constructive cases) or from Hemaspaandra, Hemaspaandra,
and Rothe~\shortcite{hem-hem-rot:j:destructive-control}
(for the destructive cases), modified in the obvious way for the nonunique-winner
case (see Observation~\ref{obs:winner-model}).~\end{proofs}

\begin{theorem}
\mcontrol{approval}{\vpaircd{}C\vpair{PC}{RPC}-\vpairtetp} are each in \p.
\end{theorem}

\begin{proofs}
Given an election $(C,V)$ and a preferred candidate of the chair $p \in C$,
we can determine in polynomial time if $p$ can be made a winner by partitioning
of candidates and by runoff partitioning of candidates as follows.
Let $k$ denote the number of manipulators in $V$.

For the constructive ``TE'' cases we do the following.
Since approval elections satisfy both WARP and unique-WARP,
we know from Theorems~\ref{thm:warp-one-part} and~\ref{thm:warp-two-part} that control is possible
if and only if control is possible using partition $(C-\{p\},\{p\})$.
Set all manipulators to approve of $p$. If that makes $p$ an overall
winner of the election, we are done. If not, let $c$ be the unique winner
of subelection $(C-\{p\},V)$ (since $p$ will participate in the runoff, %
the only way $p$ can fail to then win overall is if there is a unique 
winner of $(C-\{p\}, V)$ who beats $p$ in the
runoff). %
As just mentioned parenthetically, note 
that after manipulation, $c$'s score in this case must be greater than
$p$'s score.  If for all $d \in C - \{p,c\}$, 
$\score{c} > \score{d} + k$, $c$ will always be the unique winner of
$(C-\{p\},V)$ and so $p$ will never be an overall winner.
If there exists a candidate $d$ in $C - \{p,c\}$  such that
$\score{c} \leq \score{d} + k$, let 
$\score{c} - \score{d}$ voters approve of $d$ (in addition to $p$).
In this case, $(C-\{p\},V)$ does not have a unique winner
and so $p$ is an overall winner.

For the constructive ``TP'' cases, note that control is possible if and only
if the manipulators can vote so that $p$ becomes a winner in $(C,V)$. So the
optimal action for the manipulators is to approve of only $p$. Similarly, for
the destructive cases, control is possible if and only if the manipulators can
vote so that $p$ does not win (for the ``TP'' cases) or does not uniquely win
(for the ``TE'' cases) in $(C,V)$. So the optimal action for the manipulators
is to approve of all candidates except $p$.~\end{proofs}

\begin{theorem}
\scontrol{approval}{\vpaircd{}C\vpair{PC}{RPC}-\vpairtetp-\vpaircfmf} are each in \p.
\end{theorem}

\begin{proofs}
Given an election $(C,V)$ and a preferred candidate of the chair $p \in C$,
we can determine in polynomial time if $p$ can be made a winner by partitioning
candidates and by runoff partitioning of candidates as follows.

For the constructive ``TE'' cases,
since approval elections satisfy both WARP and unique-WARP,
we know from Theorems~\ref{thm:warp-one-part} and~\ref{thm:warp-two-part} that control is possible
if and only if control is possible using partition $(C-\{p\},\{p\})$.
The manipulators can preclude $p$ from winning if and only if there
is a candidate $c \neq p$ that can be made to uniquely win
using partition $(C-\{p\},\{p\})$.  This can easily be checked by
having all manipulators approve of only $c$.

For the constructive ``TP'' cases, note that control is possible
if and only if the manipulators cannot vote so that $p$ does not
become a winner in $(C,V)$. So the optimal action for the
manipulators, regardless of who goes first, is to approve of
all candidates except $p$. Similarly, for the destructive cases,
control is possible if and only if the manipulators cannot vote
so that $p$ becomes a winner (for the ``TP'' cases) or a
unique winner (for the ``TE'' cases) in $(C,V)$.
So the optimal action for
the manipulators, regardless of who goes first, is to approve
of only $p$.~\end{proofs}

\begin{theorem}\mcontrol{approval}{DCPV-\vpairtetp} is in \p.\end{theorem}

\begin{proofs}
Given an election $(C,V)$ and a despised candidate of the chair $p \in C$, we
can determine in polynomial time if $p$ can be precluded from winning
by partitioning voters for the ``TE'' case
as follows.

\begin{enumerate}
\item If there is a candidate, $c \neq p$ such that
   $\score{p} \le \score{c}+k$, then control is possible by
having all manipulators disapprove of only $p$ and using partition
$(V,\emptyset)$.

  \item If we are not in Case 1, the only way to preclude $p$ from being
a winner is if $p$ doesn't make it to the runoff, i.e.,  if there exist
a partition and a manipulation such that $p$ is not a unique winner
of either subelection. If in this partition we make all manipulators vote
to disapprove of only $p$, $p$ is still not a unique winner of
either subelection. So, we can check whether we are in this case by
having all manipulators vote to disapprove of only $p$, and then
using the polynomial-time algorithm for approval-DCPV-TE
from~\cite{hem-hem-rot:j:destructive-control},
modified in the obvious way for the nonunique-winner case
(see Observation~\ref{obs:winner-model}).
\end{enumerate}

For the ``TP'' case, replace ``$\leq$'' by ``$<$'' in Case 1,
and ``unique winner'' by ``winner''
and ``approval-DCPV-TE'' by ``approval-DCPV-TP'' in Case 2.~\end{proofs}

Below we state a lemma analogous to Lemma~\ref{lem:dcpv-te:plur},
but for approval elections.

\begin{lemma}\label{lem:app-dcpv-tetp}
If there exists a partition of voters such that $p$ is not an approval winner
in the ``TE'' (``TP'') model when all manipulators approve of only $p$,
then there exists a partition such that
$p$ can never be made an approval winner by the manipulators in the same tie-breaking model.
\end{lemma}

\begin{proofs}
The proof for the ``TE'' case follows similarly to the proof of Lemma~\ref{lem:dcpv-te:plur},
so we just provide the proof of the ``TP'' case.

Given an election $(C,V)$ and a candidate $p \in C$, we do the following.

Let $(V_1,V_2)$ be a partition such that $p$ is not a winner when all manipulators
approve of only $p$. If $p$ can never be made a winner by the manipulators in
this partition then we are done.
So, suppose there exists a manipulation such that $p$ is an overall winner
(with the partition $(V_1,V_2)$).
Without loss of generality $p$ is a winner of the subelection $(C,V_1)$.
Then if all manipulators in $V_1$ approve of only $p$, we know that
$p$ remains a winner of $(C,V_1)$. Note we don't get any new winners in $(C,V_1)$.
Since $p$ is not an overall winner if all manipulators approve of only
$p$ there is a candidate $c \neq p$ such that if all manipulators vote for $p$,
$c$ is a winner of $(C,V_2)$ and $\score{c} > \score{p}$.

Now move all manipulators from $V_2$ to $V_1$. Note that $c$ remains a winner
of $(C,V_2)$ and that $c$ will always beat $p$ in the runoff.
It follows that in this
new partition, $p$ is never a winner, no matter what the manipulators do.~\end{proofs}

Lemma~\ref{lem:app-dcpv-tetp} implies that %
\scontrolcf{approval}{DCPV-TE} and \scontrolcf{approval}{DCPV-TP} are both in \p,
since control is possible if and only if (nonmanipulator) control %
is possible when all manipulators approve of only $p$. This can be checked using
the corresponding polynomial-time algorithms from Hemaspaandra, Hemaspaandra,
and Rothe~\shortcite{hem-hem-rot:j:destructive-control}, modified in the obvious
way for the nonunique-winner case (see Observation~\ref{obs:winner-model}).

\begin{theorem}\label{thm:app:dcpv-cf-te-tp}
\scontrolcf{approval}{DCPV-\vpairtetp} are both in \p.
\end{theorem}

Lemma~\ref{lem:app-dcpv-tetp} above also implies that the
corresponding manipulators-first cases are both in \p. The proof of the following
theorem follows from a similar argument as the proof of Theorem~\ref{thm:plur:dcpv-mf-te}.

\begin{theorem}\label{thm:app:dcpv-mf-te-tp}
\scontrolmf{approval}{DCPV-\vpairtetp} are both in \p.
\end{theorem}

\end{document}